\documentclass[12pt]{article}

    \usepackage{newtxtext,newtxmath}
    \usepackage{bm}
    \usepackage[T1]{fontenc}
    \usepackage{physics}
    \usepackage[colorlinks=true, linkcolor=blue, citecolor=blue, breaklinks=true, urlcolor=blue]{hyperref}
    
    \usepackage{graphicx}
    \usepackage[dvipsnames]{xcolor}
    \usepackage[letterpaper,margin=1in]{geometry}
    
    \renewenvironment{abstract}
    	{\quotation}
    	{\endquotation}
    
    \date{}
    
    \makeatletter
    \renewcommand{\fnum@figure}{\textbf{Figure \thefigure}}
    \renewcommand{\fnum@table}{\textbf{Table \thetable}}
    \makeatother
    
    \newcommand{\R}{\text{Re}}

    \usepackage{cite} 

    \usepackage{url}
    
    \def\scititle{
    	Surfing on metachronal waves: ciliary transport by inertial coasting
    }
    
    \title{\bfseries \boldmath \scititle}

\author{
    Rafał Błaszkiewicz$^{1,2\dagger}$,
    Margot Young$^{1\dagger}$,
    Albane Th{\'e}ry$^{1,3,4\ast}$, 
    Talia Calazans$^{1}$, \and
    Yoichiro Mori$^{4}$,
    Maciej Lisicki$^{1,2\ast}$, and
	Arnold J. T. M. Mathijssen$^{1\ast}$. \and
    \small$^{1}$ Department of Physics and Astronomy, University of Pennsylvania, Philadelphia, USA
    \and
    \small$^{2}$ Faculty of Physics, University of Warsaw, Warsaw, Poland	
    \and
    \small$^{3}$ The Zeeman Institute for Systems Biology and Infectious Disease Epidemiology Research, \and \small School of Life Sciences and Mathematics Institute, University of Warwick, Coventry, UK
    \and
    \small$^{4}$ Department of Mathematics, University of Pennsylvania, Philadelphia, USA
    \and
	{\small {$^\ast$Corresponding authors. albane.thery@warwick.ac.uk, mklis@fuw.edu.pl, amaths@upenn.edu.}}
    \and
	\small$^\dagger$These authors contributed equally to this work.
}

\begin{document} 

\maketitle

\begin{abstract}
Motile cilia drive biological fluid transport through whip-like beating motions that synchronize into metachronal waves. The lengths of these cilia span three orders of magnitude, from microns in human airways to millimeters in ctenophores. While recent studies have considered ciliary flows at intermediate Reynolds numbers, the effect of inertia on coordinated particle transport remains unexplored. Here, we address this gap using “Pufflets,” the inertial counterparts of Stokeslets. These Pufflets describe rapidly accelerating flows generated by short-lived impulses, encoded by spatiotemporally singular momentum injections. To produce such rapid impulses experimentally, we designed an Atwood machine that generates long-lived Pufflet flows, which we capture with high-speed PIV measurements that agree well with analytical theory and simulations. 
Moreover, we find that pairs of equal and opposite Pufflets can drive net particle displacements and mixing due to time reversal symmetry breaking, which would be impossible in Stokes flow.
Finally, we consider metachronal waves of Pufflets. Remarkably, we discover that particles can surf on these waves by coasting inertially from one cilium to the next, leading to highly efficient particle transport. This work paves the way toward understanding rapidly accelerating flows and collective transport driven by biological and artificial cilia.

\end{abstract}


\section{Introduction}
\label{sec:Introduction}

\begin{figure}[ht!]
 	\centering
\includegraphics[width=\textwidth]{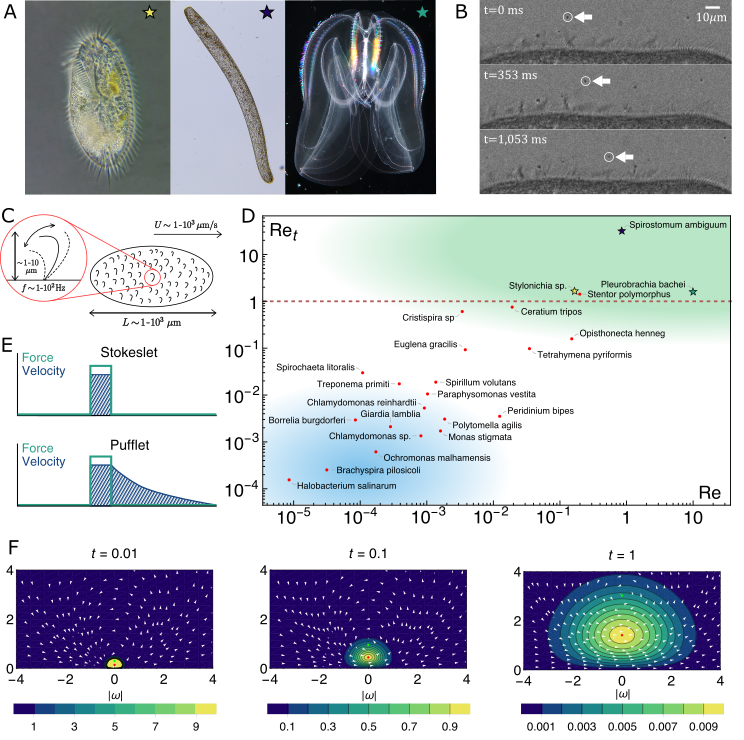}
 	\caption{\textbf{Ciliary transport across the scales.} 
    \textbf{(A)} Organisms using cilia at intermediate transient Reynolds numbers: \textit{Stylonichia spp.} (left), \textit{Spirostomum ambiguum} (center) and the ctenophore \textit{Mnemiopsis leidyi} (right). Image by Bruno Vellutini, CC3.0.
    \textbf{(B)} Tracer particle transported by a metachronal wave on the cell surface of \textit{Spirostomum ambiguum}.
    \textbf{(C)} Characteristic length and time scales determining the Reynolds number $\R$ and transient Reynolds number $\R_t$.
    \textbf{(D)} Map of $\R$ and $\R_t$ for ciliated organisms across the scales. From databases \cite{velho_rodrigues_bank_2021} and \cite{heimbichner_goebel_scaling_2020}. 
    \textbf{(E)} Flow relaxation for non-inertial Stokeslets and inertial Pufflets. 
    \textbf{(F)} Diffusion of vorticity generated by a Pufflet. Contour plots present the magnitude of vorticity. The point of maximal vorticity (red dot) does not coincide with the vortex center (green dot) surrounded by streamlines (white).
    }
    \label{fig1}
\end{figure}

\noindent
Cilia are slender motile organelles used for propulsion by a remarkably diverse range of eukaryotic organisms, from unicellular microswimmers to centimeter-sized comb jellyfish (Fig.~\ref{fig1}A) \cite{Brennen1977FluidFlagella, multiscalephys, Lauga2009TheMicroorganisms, 10.1093/icb/icab179}.  Through their rhythmic beating, cilia generate fluid motion that enables locomotion, transport of nutrients \cite{10.1038/s41598-020-64420-7}, streaming, and coordinated biological stimulus responses \cite{PhysRevLett.96.058102, 10.1126/science.aae0450, SpontaneousCreation}.
When multiple cilia operate together, they often synchronise to form metachronal waves:  travelling patterns that enhance directed fluid transport and particle entrainment along the ciliated surfaces (Fig.~\ref{fig1}B) \cite{doi:10.1073/pnas.1218869110, doi:10.1073/pnas.2214413119, Solovev_2022, Byron2021MetachronalDirections}. Such transport is vital for the survival and well-being of organisms across many scales, including in humans, where motile cilia in the sinuses and lungs 
\cite{Nawroth2020MultiscaleLung,Ramirez-SanJuan2020Multi-scaleArrays,Sears2013}, brain \cite{Worthington1963}, and fallopian tubes \cite{Lyons2006} drive critical clearance and guidance mechanisms. Cilia-related dysfunctions of these systems––ciliopathies––underlie serious multi-systemic diseases 
\cite{ciliarelateddiseasesrev}.

Inspired by the versatility and efficiency of biological cilia, numerous artificial ciliary systems which exhibit synchronization \cite{Bruot2016RealizingColloids} and metachronal motion \cite{biomimetics9040198,doi:10.1126/sciadv.abd2508} have been developed for applications that include droplet\cite{Demirors2021AmphibiousCarpets, doi:10.1021/acsami.0c10034} and particle\cite{https://doi.org/10.1002/adfm.201706666, benBioinspiredMagneticResponsive2020} transport, microrobot locomotion\cite{doi:10.1021/acsami.1c03009, guMagneticCiliaCarpets2020a}, and small-scale mixing\cite{D1SM01680F}. In contrast to their biological counterparts, however, artificial and engineered cilia are often larger, stiffer, or driven externally at higher frequencies than the natural beating \cite{ulIslam2022MicroscopicReview}. Such conditions can amplify inertial effects, which brings into question the validity of descriptions based on purely viscous dynamics, which are commonly used to model microscale flows. 

Despite extensive progress, a complete model of ciliary flows remains elusive due to their multiscale nature. Most hydrodynamic models rely on the assumption of viscosity-dominated dynamics and employ the classical Stokesian framework. However, this approach completely neglects inertial phenomena, particularly those associated with fast actuation and the resulting diffusion of momentum. 
Cilia usually beat tens of times per second, so their rapid accelerations can excite an inertial response in the fluid. For example, organisms including {\it Pleurobrachia} ctenophores \cite{heimbichner_goebel_scaling_2020}, {\it Spirostomum} ciliates \cite{Mathijssen2019}, and {\it Chlamydomonas} algae \cite{Wei_Dehnavi_Aubin-Tam_Tam_2021} clearly show inertial effects. 

To quantify the importance of these inertial effects, we describe the flow characteristics in terms of the organism length $L$, velocity $U$, ciliary beat frequency $f$, and the fluid kinematic viscosity $\nu$ using two dimensionless parameters (Fig.~\ref{fig1}C).
First is the Reynolds number, $\R := UL/\nu$, which compares advective and diffusive transport of momentum in the fluid.
Second is the transient Reynolds number $\R_t := L^2/\nu T_\textmd{acc}$, 
which compares the time scale of actuation $T_\textmd{acc}=1/f$ to that of momentum diffusion, $\tau_d = L^2/\nu$.
Large $\R_t$ numbers correspond to rapid accelerations.

In Figure \ref{fig1}D, we map out this $(\R,\R_t)$ space based on available literature data for representative ciliated organisms \cite{velho_rodrigues_bank_2021}. Most bacteria and small ciliates occupy the stationary Stokesian regime of $\R,\R_t \ll 1$, where inertial effects of any kind can be safely neglected. However, larger and more rapidly actuated systems such as {\it Pleurobrachia} and {\it Spirostomum} (Fig.~\ref{fig1}A,D; colored stars) lie within the transitional regime where $\R\ll 1$ but $\R_t\sim 1$. In this regime, momentum transport remains diffusive but has time scales comparable to those of system actuation, calling for models that resolve the resulting unsteady flow fields. 

Previous studies of this transitional regime \cite{PhysRevLett.122.124502, PhysRevResearch.7.L012029, 10.1115/IMECE2023-112572, PhysRevFluids.5.063103, 10.1093/icb/icab179, PhysRevFluids.6.053102, C4SM00770K} indicated that fluid inertia can influence ciliary cooperation \cite{PhysRevResearch.7.L012029,Herrera-Amaya_2024} and metachronal wave formation \cite{PhysRevResearch.7.L012029,10.1115/IMECE2023-112572,10.1093/icb/icab179}, but it can also affect fluid transport and mixing \cite{PhysRevFluids.6.053102,
Zhou2017,intReSquirmer,Ryu2010,Nagai2013,B901660K}, and even promote coordinated motion in highly viscous environments \cite{PhysRevResearch.7.L012029,C4SM00770K,Winkler2016LowSimulations}. Nevertheless, it remains unclear how fluid inertial effects alter directed ciliary transport. In particular, the role of inertial coasting––the persistent motion of suspended particles following rapid actuation––has not been explored systematically. 

In this paper, we address this gap by modeling ciliary flows as short but powerful impulses of force, with small $\R$ but large $\R_t$, which we call ``Pufflets.'' 
We study these Pufflet flows theoretically and experimentally with a macroscale physical model using high-speed flow field imaging. 
After elucidating the properties of a single Pufflet, we study the effect of two opposing Pufflets that approximate the power and recovery stroke pattern of biological cilia. 
Finally, we examine waves of Pufflets as a basic model for inertial metachronal transport. 

\section*{Results}

\subsection*{Pufflets: impulsive flows at intermediate transient Reynolds number}

Formally, the Pufflet corresponds to the Green's function of the linearized Navier-Stokes equations, also known as the unsteady Stokes equations, and provides a minimal model for fast fluid actuation \cite{Pozrikidis1989}.
It can be regarded as the unsteady (i.e., stemming from a time-dependent equation) analogue of the classical Stokeslet \cite{Lauga2009TheMicroorganisms}. 
In the steady case, fluid motion occurs only while the Stokeslet force is applied; once the force ceases, the flow instantly stops. 
In contrast, the Pufflet injects momentum into the fluid, generating a transient flow that persists after the cessation of the force and decays over time. 
A schematic representation of this concept is shown in Fig.~\ref{fig1}E.

This simple construct is a useful approximation to describe inertial flows. 
In general, solutions to the unsteady Stokes equations can be constructed by time-integrating forces over the memory kernel of the Green's function (see Methods).
Introducing Pufflets simplifies this formalism by approximating the force as a Dirac delta function in time, making the flows analytically tractable and providing a valuable building block for reduced models of inertial microhydrodynamics. 

The Pufflet flow has the same spatial symmetries as the steady Stokeslet field, but the amplitudes of the diagonal and off-diagonal terms decay with time. 
The flow evolution is expressed as a function of a non-dimensional variable, $r/\sqrt{\nu t}$, which represents a natural comparison between the spatial position and the length scale of viscous vorticity diffusion at any given time: initially, vorticity is generated locally where the force is applied, and then it diffuses outward to form a ring-like vortex. As the vortex center travels away from the origin in the direction perpendicular to the applied impulse, its strength gradually decays. 

Interestingly, as shown in Fig.~\ref{fig1}F, the location of the maximum vorticity (red point) does not coincide with the center of the vortex (green point), defined as the location where the flow velocity vanishes.
While the latter can only be obtained numerically as $r_\textmd{vortex}\approx 3.022 \sqrt{\nu t}$, in the direction perpendicular to the force, the maximum vorticity location can be derived analytically: $r_\omega=\sqrt{2\nu t}$ (see Methods).

\subsection*{High-speed impulse PIV experiments validate theory}

To test these theoretical predictions, we designed an experimental setup capable of exploring both steady (Stokesian) and unsteady (inertial) flow regimes on the macroscale. 
The system consists of a sphere that is rapidly accelerated in a viscous fluid using an Atwood machine (Fig.~\ref{fig2}A-B).  
The sphere in the fluid (shown as a black disk) is connected using a thin string to a mass outside the tank (blue square). 
First, this mass is lifted up and released, such that it accelerates freely under gravity while the string is slack. 
Then, as the string becomes taut, the falling mass delivers a sharp pull on the sphere in the fluid. 
However, this force is very short-lived, about a few milliseconds, and then the mass hits the base of the setup and comes to rest. 
Thus, the force approximates a strong but infinitesimally short impulse, representative of a Pufflet.

The resulting flow fields are visualized with high-speed Particle Image Velocimetry (PIV) using tracer particles illuminated by a planar laser sheet (see Methods). The apparatus enables repeatable experiments, and it allows for independent control of both the Reynolds number ($\R$) and the transient Reynolds number ($\R_t$) by tuning the mass and the impulse duration.

The inertial character of these Pufflet flows can be seen by comparing the velocity of the sphere with that of the surrounding liquid (measured by PIV). Although the two profiles in Fig.~\ref{fig2}C share the same overall shape, they are shifted in time:
the liquid velocity lags behind the sphere motion by a delay corresponding to the viscous diffusion timescale, $\tau$.
Comparison between the experimentally obtained velocity field and the theoretical Pufflet model shows excellent agreement (Fig.~\ref{fig2}D). 

To probe the comparison further, we identify the vortex position in both experiment and theory (Fig.~\ref{fig2}E), again finding very good correspondence. Repeating the vortex location analysis across successive time frames allows us to track the vortex motion in time (Fig.~\ref{fig2}F). The observed diffusion of vorticity closely follows the theoretically predicted scaling of $r_\textmd{vortex}\sim3.022\sqrt{\nu t}$ (described above) in Fig.~\ref{fig1}F.

The pattern of evolving vorticity motivates examination of the Lagrangian transport generated by a Pufflet. As shown in Fig.~\ref{fig2}G, the deformation of an initially uniform grid of fluid parcels reveals several interesting features. Some particle trajectories form transient loops: the vorticity diffusing outward momentarily traps tracers before releasing and propelling them forward. 
The flow front assumes a characteristic ``mushroom cap'' shape, enclosing a region of well-mixed fluid. These observations motivated us to explore how such impulsive flows can drive mixing when subjected to periodic forcing.

\subsection*{Cyclet: mixing by equal and opposite impulses} 

To investigate mixing induced by rapid impulses, we introduce the Cyclet: a pair of oppositely directed Pufflets applied sequentially at the same location with a controlled time delay. Experimentally, this is achieved by adding a second Atwood machine to the setup (Fig.~\ref{fig3}A), allowing upward and downward impulses to be delivered in succession.

We first tracked individual tracer particles through the complete Cyclet motion and compared the experimental trajectories with theoretical predictions initiated from the same starting positions (Fig.~\ref{fig3}B). Once again, excellent agreement was found. 
Moreover, the forward trajectories do not coincide with the backward trajectories, leading to loop-like structures. 
This demonstrates a clear breakdown of time reversibility, a hallmark difference between inertial and Stokesian flows.

Moreover, the loops are not closed: their final positions differ from the initial position, unlike the famous video by G.I.~Taylor \cite{taylor1967video}.  
Figure~\ref{fig3}C shows histograms of tracer displacements (final minus initial positions, normalized by the displacement after the one-way impulse) for both cases. While the non-inertial back-and-forth Stokeslets produce no net displacement (within experimental uncertainty), the Cyclet results in significant tracer displacements. 

Repeating the Lagrangian visualization from Fig.~\ref{fig2}G for the Cyclet (Fig.~\ref{fig3}D) reveals a complete break-down of time reversal symmetry: trajectories are clearly non-closed, and two well-mixed regions form at the edges of the “mushroom caps.” Enhanced vertical transport can also be seen as the Lagrangian grid develops a trough-like deformation.

To measure mixing in this system, we consider two equal sets of tracer points initially placed in adjacent subdomains (green and blue) and track their evolution over successive Cyclet iterations (Fig.~\ref{fig3}E). After the first cycle, the mixed region is small and localized near the Pufflet fronts. With each iteration, the mixed area expands, accompanied by a gradual entrainment of external fluid.

To quantify the degree of mixing, we define the mixing number (following Ref.~\cite{ding_mixing_2014}; see Methods) as the mean minimal distance between the points from the two tracer sets. The evolution of the mixing number, scaled by the number of iterations of the Cyclet, is shown in Fig.~\ref{fig3}F. At longer times, the scaled mixing number slowly decreases, suggesting that the mixing is not chaotic. This is consistent with expectations for mixing driven by spatially focused forcing in an unbounded viscous domain \cite{Aref1986}. In summary, the Cyclet shows the benefit of moving from a purely Stokesian regime to the inertial one by unlocking net transport and efficient mixing, thus circumventing the scallop theorem.

\subsection*{Long-ranged particle transport by surfing on a wave of Pufflets}

After considering single Pufflets and Cyclets, we turn our attention to particle transport by metachronal waves of Pufflets as a model for rapidly accelerating cilia, driving flows at intermediate $\R_t$. 
As sketched in Fig.~\ref{fig4}, we construct a line of Pufflets that are fired one after another to generate a traveling wave. 
The essential question is that of cooperation: when does this wave give rise to emergent large-scale transport? 

To analyze this systematically, we non-dimensionalize the problem and set the Pufflet force strength to 1.
For simplicity, we consider the regime where the time interval $\Delta t$ between successive Pufflets is large compared to the inertial time scale, allowing the flows to settle before the next Pufflet fires.
Then, the only relevant parameter is the non-dimensional spatial separation $\delta$. 
In Fig.~\ref{fig4}A, we plot the deformation of a fluid parcel grid by such a line of firing Pufflets for decreasing $\delta$ values. 
We identify three distinct regimes: (i) distant Pufflets, (ii) cooperation, and (iii) transport. 

In the case of large separations (i), spatially distant Pufflets interact only weakly, as shown for $\delta = 1$. In particular, the front of a given Pufflet mushroom cap does not reach the next Pufflet (red point). 
In this regime, no long-ranged transport occurs: if we plot the trajectories of particles in the fluid (Fig.~\ref{fig4}B; blue tracks), the wave only transports particles over finite distances less than $2 \delta$. 

In the cooperative regime (ii), at a critical separation distance, $\delta \sim 0.5$, the front of the mushroom cap from the first Pufflet is also advected by the second Pufflet. 
The continued transport of this front results in more complex grid deformations, characterized by local shedding and stretching of successive vortex structures. 
In this intermediate regime, particles starting right in front of the wave are transported forward over large distances (yellow trajectories). 
However, off-axis particles eventually leave the moving front: they can only be transported by a few Pufflet separations. 

Finally, in the transport regime (iii), for Pufflets even closer to one another, a structure forms at the beginning of the wave and is transported along it, as plotted for $\delta = 0.25$ in Fig.~\ref{fig4}A. Particles located near the origin of the wave are then transported in a flow region that is consistently folded by each successive Pufflet. 

The trajectories of the transported particles are particularly complex, and we analyze them in more detail by considering Poincaré maps in the reference frame of the wave (Fig.~\ref{fig4}C). To compute them, we adjust the origin at each firing event to the location of the last active Pufflet. 
Each line of a given color in the map represents one of the transported particles, with each dot showing its position at subsequent Pufflet actuation. 
Tracer points get transported around the folded structure, with significant jumps occurring in a region near the Pufflet. 
This creates complex, non-periodic trajectories over many Pufflets. In particular, the transported liquid is well mixed, consistent with our study of the Cyclet above. 

Finally, we compute the three-dimensional volume of the liquid displaced by a wave of Pufflets (Fig.~\ref{fig4}D). 
This volume is large for small separation distances, e.g.\ $\delta=0.15$, but significantly smaller for $\delta=0.25$.
As we approach the critical separation distance, the number of particles that are carried by the wave tends to zero, as expected.
The transported volume scales as $\delta^{-3}$ (dashed line), which we explain with an analytical scaling analysis in the Supplementary Material.

In summary, below a critical Pufflet separation, particles are capable of surfing on metachronal waves. Despite the typical limitation of low Reynolds number dynamics, here the particles can exploit inertia by coasting from one Pufflet to the next.
As such, the velocity of these particles equals the wave speed itself, which is much larger than the average velocity of the surrounding fluid. Therefore, this mechanism enables highly efficient and directed transport that can be harnessed for a wide range of applications, as discussed below.

\section*{Discussion}

In this paper, we study the transport driven by metachronal waves of rapidly accelerating cilia. 
To model the flows generated by these cilia at intermediate transient Reynolds numbers, we employ Pufflets: Green's functions of the linearized Navier-Stokes (LNS) equations that describe an infinitesimally short impulse of force applied at a single point.
To create these Pufflets, we construct an experimental Atwood system capable of delivering rapid kicks at $\R_t >1$ values while keeping $\R < 1$. The setup uses a highly viscous fluid to maintain these conditions at the macroscopic scale, allowing for the probing of detailed dynamics at high resolution.
We use high-speed PIV experiments to measure the Pufflet flows and vortex dynamics, and compare these with the theoretical predictions, finding good agreement. 
To emulate the power and recovery strokes of biological cilia, we also investigate Cyclets: two Pufflets separated in time that fire in opposite directions. 
We find that Cyclets effectively break time reversibility, leading to mixing and net transport despite complete spatial symmetry, circumventing the constraints of Purcell's Scallop Theorem \cite{Hubert2021ScallopMesoscale}. 
Finally, we study metachronal waves of Pufflets.
Particles can surf on these waves using inertial coasting, enabling rapid and long-range transport:
In Stokes flow, particles need to be directly handed off from one cilium to another, or they require multiple beat strokes to bridge distances larger than a cilium length.
With inertia, they continue moving after the ciliary force has stopped, allowing particles to move between distant cilia with a single beat stroke.
Together, our findings establish inertial coasting as a new fundamental mechanism for transport.

This concept of inertial coasting can be investigated more extensively. 
For simplicity, we considered waves of Pufflets separated by time intervals $\Delta t$ much larger than the inertial relaxation time $\tau$.
Removing this constraint introduces a new dimensionless `relaxation' number, $\text{Rx} = \Delta t/\tau$. 
The wave moves slowly for large Rx values, reducing the long-range transport speed.
At small Rx numbers, however, the wavefront could move too quickly for the particles to catch up, reducing the total transported volume.
At an optimal intermediate Rx, inertial coasting could resonate with the Pufflet driving frequency, leading to optimal transport. 
In future work, connecting this framework with other applications in inertial microfluidics \cite{DiCarlo2009InertialMicrofluidics} for particle separation, droplet micromanipulation, cellular sample processing, and biomedical diagnostics \cite{Zhang2016FundamentalsReview} would be of great interest.

Moreover, we consider Pufflets in the absence of surfaces because the Green's function \cite{pozrikidis_introduction_1997, Winkler2016LowSimulations} of the linearized Navier-Stokes equation in half-space with no-slip conditions have only been studied in the frequency domain and cannot be expressed in closed form \cite{Fouxon_2018}. 
For water-air interfaces, it is possible to use the method of images \cite{Mathijssen2019}, but determining the flow close to solid surfaces remains an exciting challenge for future research, especially in the context of ciliated epithelia. 
Regardless of surfaces, we expect that the main physical principles for surfing on metachronal waves will be similar. 

After analyzing particle surfing with simulations and theory, the next step is to build metachronal waves in experiments. Recent articles have started exploring artificial cilia at intermediate $\R$ in soft-robotic platforms  
\cite{Ford2019HydrodynamicsLag, Peterman2024EncodingPlatform}, but rapid accelerations with intermediate $\R_t$ remain underexplored.
A first challenge is to nanofabricate active Cyclets to optimize mixing in microfluidic devices \cite{Toonder2008ArtificialMixing}.
The next challenge would be to construct waves of Pufflets. 
One way to realize this could be to fabricate a series of miniaturized mouse traps that fire sequentially, with precise control over the spatial separation $\delta$ and the time interval $\Delta t$. This could lead to the development of highly efficient fluidic conveyor belt systems where cargo is transported with optimal inertial coasting.  
More generally, in a broad range of engineered and biological systems, it is of critical importance to understand rapid accelerations \cite{Ilton2018TheSystems}.
Pufflets could help make these systems more analytically tractable.


\subsection*{Acknowledgments:} 
A.J.T.M.M. acknowledges funding from the Charles E. Kaufman Foundation (Early Investigator Research Award, KA2022-129523; and New Initiative Research Award KA2024-144001), the National Science Foundation (UPenn MRSEC, DMR-2309043), the University of Pennsylvania (CURF, VIPER, Vagelos MLS, and FERBS programs), and the Research Corporation for Science Advancement (Cottrell Scholar Award CS-CSA-2026-125).
The work of R.B. and M.L. was supported by the National Science Centre of Poland Sonata Bis grant no. 2023/50/E/ST3/00465 to M.L. 
M.L. is grateful to the Fulbright Commission for the  Fulbright Senior Award to spend a sabbatical in the Mathijssen lab. 
Y.M. was partially supported by the NSF (Grant DMR2309034), and the Math+X award from the Simons Foundation (Award ID 234606).
M.Y. is supported by the NSF Graduate Research Fellowship Program (Grant DGE-2236662). Any opinions, findings, and conclusions or recommendations expressed in this material are those of the author(s) and do not necessarily reflect the views of the NSF. 
We acknowledge all lab members for helpful discussions and critical feedback.

% ...

\subsection*{Competing interests:}
There are no competing interests to declare.

\subsection*{Data and materials availability:}
\subsection*{Supplementary materials}
Materials and Methods\\
Supplementary Text\\
Figs. S1 to S2\\
References \textit{(7-\arabic{enumiv})}\\ 
Movies S1 to S6\\


\clearpage

\section*{Materials and methods}

\subsection*{Pufflet - inertial Stokes flow singularity}
\label{Pufflet}
In this paper, we investigate inertial effects at intermediate transient Reynolds numbers. 
This regime is defined using the incompressible non-dimensional Navier-Stokes equations, which can be written in the dimensionless form using $L$ as the typical length scale and $U$ as the characteristic velocity as
\begin{equation}
\R_t
 \frac{\partial \tilde{\bm{u}}}{\partial \tilde{t}}  + \R ~  (\tilde{\bm{u}}\cdot \nabla)\tilde{\bm{u}}=-\nabla \tilde{p} +  \nabla ^2 \tilde{\bm{u}} + \tilde{\bm{f}}, \qquad \nabla \cdot  \tilde{\bm{u}} = 0
 \label{navstokesnondim}
\end{equation}
where $p$ is pressure, $f$ is the force applied to the fluid, and tildes denote rescaled variables. There are two dimensionless parameters that determine the flow: the Reynolds number ($\R$) and the transient Reynolds number ($\R_t$). Assuming a separation of times scales between that of the fluid actuation ($T_\text{acc}$) and the momentum diffusion ($L_0^2/\nu$) where $\nu$ is the kinematic viscosity, we can write them as
\begin{equation}
\R=\frac{U L}{\nu} \qquad \text{and} \qquad \R_t=\frac{L^2}{\nu T_{\text{acc}}},
\label{ReDefs}
\end{equation}
where $\nu$ is kinematic viscosity, and $T_{\text{acc}}$ is time scale of acceleration. For typical microswimmers and dynamics in microfluidic systems, when the Reynolds number is low, so is the transient Reynolds number \cite{velho_rodrigues_bank_2021,Lauga2009TheMicroorganisms}, simplifying the Navier-Stokes equations to the Stokes equations\cite{pozrikidis_introduction_1997}. 
However, for rapid accelerations (i.e., small $T_{\text{acc}}$) or spatially extended systems, the transient Reynolds number $\R_t$ can be order 1 while $\R\ll1$. The Navier-Stokes equations then reduce to the linearized Navier-Stokes equations (LNS), given by
\begin{equation}
    \rho \frac{\partial \bm{u}}{\partial t} =-\nabla p + \mu \nabla ^2 \bm{u} + \bm{f}, \qquad \nabla \cdot  \bm{u} = 0.
    \label{timedepstokes}
\end{equation}

Thanks to the linearity of LNS, the general solution to Eq. \eqref{timedepstokes} takes the form \cite{pozrikidis_introduction_1997}
\begin{equation}
\bm{u}(\bm{r},t)=\int d^3\bm{r}' \int_0^t dt' \bm{Q}(\bm{r}-\bm{r'},t-t')\cdot \bm{F}(\bm{r'},t'),
   \label{solution}
\end{equation}
where $\bm{F}(\bm{r},t)$ is the force density at position $\bm{r}$ and time $t$, and $Q(\bm{r},t)$ is Green's tensor
\begin{equation}
        \bm{Q}_{ij}=\frac{1}{\rho} \left(A(\bm{r},t)\delta_{ij} -B(\bm{r},t) \frac{\bm{r}_i \bm{r}_j}{r^2}\right),
\end{equation}
which shares the tensorial structure with the classical Stokeslet solution, albeit with time- and space-dependent amplitudes given by
\begin{align}
    A(\bm{r},t)&= \left( 1+ \frac{2 \nu t}{r^2}\right) \alpha(\bm{r},t)-\frac{\beta(\bm{r},t)}{r^2}, \\
    B(\bm{r},t)&=\left( 1+ \frac{6 \nu t}{r^2}\right) \alpha(\bm{r},t)-\frac{3\beta(\bm{r},t)}{r^2}, \\
    \alpha(\bm{r},t)&=\frac{1}{(4\pi \nu t)^{3/2}} \exp\left(-\frac{r^2}{4 \nu t}\right), \\
\beta (\bm{r},t)&=\frac{1}{4 \pi r} \text{erf} \left(\frac{r}{\sqrt{4 \nu t}}\right).
\end{align}
We note here that the amplitudes do not depend on the spatial and temporal variables independently, but rather on the combination $r/\sqrt{\nu t}$ that compares the distance with the fluid momentum diffusion length.

The simplest case for which Eq. \eqref{solution} evaluates analytically is taking a localized force impulse $ F(\bm{r},t)= f_0 \bm{e}_i \delta(t)\delta(\bm{r})$ applied at the origin, where $f_0$ is the amplitude of the impulse (with the units of momentum, Ns), $\delta$ is the Dirac delta function and $\bm{e}_i$ is the unit vector in the $i$ direction. Assuming such a form of the driving force, the integrals in Eq. \eqref{solution} are straightforward and equal to the value of the integrand at $\bm{r'}=0$ and $t'=0$. This case can be seen as the flow singularity corresponding to the Stokeslet, but for the unsteady Stokes flow (the Pufflet). It can be understood as the simplest mathematical model of a rapid impulse of force exerted on a viscous fluid. The Pufflet flow field decays rapidly with time, shedding a vortex ring that progressively broadens and becomes weaker, as shown in Fig.~\ref{fig1}F. It is worth noticing that the center of the vortex is at a different location than the maximum of the vorticity, which, for the case of a Pufflet acting along the $z$ axis, is given in cylindrical coordinates $(\varrho,z)$ by 
\begin{equation}
    |\omega|= \frac{\varrho}{16 \pi^{2/3} t^{5/2}} e^{-\frac{\varrho^2+z^2}{4t}}
\end{equation}
and its maximum yields $\varrho_{\text{max}}=\sqrt{2\nu t}$, which differs from the numerically obtained root of the velocity profile at the axis perpendicular to the Pufflet axis, $\varrho_0 \approx 3.022 \sqrt{\nu t}$.

\subsection*{Numerical methods}  
Although the Pufflet flow field is known analytically, the trajectories of the tracer points in this flow field can be obtained only numerically. It poses a question of proper numerical integration of the flow field, whose magnitude is initially very high in a small region, but later decays by orders of magnitude while spreading over a larger region. Because this system spans multiple scales, we must introduce a different time variable $s$ 
\begin{equation}
    \frac{d \mathbf{r}}{ d s } = \frac{d \mathbf{r}}{ d t }\frac{d t}{ d s } =  \bm{u}_\text{puff}(\bm{r},t) \frac{d t}{ d s }.
\end{equation}
We choose $s$ so that 
\begin{equation}
    \frac{d t}{ d s } =  |\bm{r}|^3 ,
\end{equation}
and obtain the system of equations
\begin{equation}
    \left\{ \begin{aligned}
        \frac{d \bm{r}}{ d s } &=  |\bm{r}|^3 \bm{u}_\text{puff}(\bm{r},t)  \\
         \frac{d t}{ d s } &= |\bm{r}|^3, 
    \end{aligned} \right.
\end{equation}
which can be integrated robustly with traditional numerical schemes. In this work, we use the \texttt{NDSolve} function of Wolfram Mathematica.

\subsection*{Mixing number}
To quantify the degree of mixing, we employ a measure called the mixing number \cite{ding_mixing_2014}. Its construction is based on the evaluation of distances between points originating from two distinct fluid subdomains. In our case, we consider two adjacent square-shaped regions (located to the left and right of the Cyclet - see Fig.~\ref{fig3}E), each containing $N=10^4$ points arranged on a regular grid. The mixing number is defined as
\begin{equation}
m=\left( \prod^{N}_{i=1} \min_{j\in \{1,..,N\}} |\mathbf{a}_i - \mathbf{b}_j|\right)^{1/N},
\end{equation}
where $\mathbf{a}_i$ denotes the position of $i$-th point in the first set, and $\mathbf{b}_j$ is the position of $j$-th point in the second set. The indices $i,j\in \{1,..,N\}$ iterate over the two respective point sets. 

The mixing number can be interpreted as the geometric mean of the minimal distances between points belonging to different sets (or colors). This measurement is not symmetric under the exchange of the two sets and depends on the extent and location of the chosen subdomains—a limitation shared by most metrics attempting to quantify mixing \cite{ding_mixing_2014, krasnopolskaya_mixing_1999,stone_imaging_2005, mathew_multiscale_2005, reece_new_2020}. The mixing number can be evaluated at each time step during the mixing process; in the case of the Cyclet, it can be calculated after each iteration. To facilitate comparison, we normalize the mixing number by its initial value $m_0$, obtained at the beginning of the process. Assuming the exponential dependence
\begin{equation}
    m=m_0\exp(-\eta_m N),
\end{equation}
where $\eta_m$ denotes the mixing rate, we can analyze how $\eta_m$ evolves with the number of iterations. If $\eta_m$ were constant, the mixing process would be purely chaotic, which can be understood based on Lyapunov exponents \cite{lyapunov}. In chaotic flow, if a control volume of fluid is being stretched by a factor $e^{\lambda t}$ (where $\lambda$ is the Lyapunov exponent) in one direction, then from volume conservation it is being compressed by  $e^{-\lambda t}$ in other direction, hence bringing every particle of that volume closer to the exterior particles.  \\
However, as shown in Fig.~\ref{fig3}F, the mixing rate decreases over time. This observation indicates that the mixing slows down relative to the chaotic regime, which is expected, since the domain is unbounded and sample points gradually drift away from the region where the mixing is driven. Nevertheless, the mixing induced by the Cyclet remains significant.
\subsection*{Experimental Methods}
\label{experimentmethods}

To experimentally investigate Pufflets and Stokeslets, we study the motion of a sphere inside a tank of liquid (Fig.~\ref{fig2}A). The tank is made of clear acrylic ($20\times20\times40$~cm) and filled with high viscosity silicone oil (Clearco PSF-10,000~cSt Pure Silicone Fluid) seeded with 10~$\mu$m silver-coated hollow glass tracer particles (Dantec Dynamics S-HGS-10). The sphere has a 12.7~mm diameter and is painted black to mitigate reflections. We actuate the sphere using a 0.6~mm diameter extra-flexible, low-stretch braided wire (McMaster-Carr) attached to an Atwood machine (Eisco Labs) and a hanging mass. We illuminate a plane of tracer particles for flow visualization using a 1W CW 520~nm laser sheet (CivilLaser). A high-speed camera (Phantom v1840) perpendicular to the laser sheet is used to image the illuminated plane at 1500~fps. 

To experimentally achieve a single Pufflet, a sharp impulse must be delivered to the sphere inside the tank. To deliver this impulse, we use the process shown in Figure~\ref{fig2}B. First, a mass of approximately 60~g is released by an electromagnet and allowed to fall freely with no tension, so it accelerates quickly. Then the string becomes taut and the mass falls under tension for the last 1~cm (approximately), applying an upward force to the sphere. The mass is caught at the bottom of its trajectory by a magnet, which prevents bounces and stops the motion of the mass, so the sphere comes to rest. To ensure that the sphere stops quickly, it is also tethered to the bottom of the tank by another string. 

This system allows full control of flow conditions and the possibility of realizing multiple flows, varying $\R$ and $\R_t$ almost independently. By delivering a sharp impulse to the sphere, we obtain a flow with intermediate $\R_t$, while keeping $\R$ low. When a lower $\R_t$ is required, the sphere is actuated by manually pulling on the string instead of using a falling mass. The Pufflet shown in Figure~\ref{fig2}C-F has $\R = 0.49$ and $\R_t = 2.19$, and a more detailed explanation of this calculation can be found in the Supplementary Information. 

For each video, the velocity fields were found with Particle Image Velocimetry (PIV) using PIVlab software \cite{Thielicke_2021}. Additional analysis and visualization, including the construction of streamlines (Fig.~\ref{fig2}D), were completed in Python. For the single Pufflet case, three flow fields from separate trials were averaged to give the flow field used in Figure~\ref{fig2}C-F. 

In particular, we tracked $v_y$, the component of velocity in the direction of the sphere's motion, at a fixed point in the liquid throughout the experiment to compare the inertia of the sphere and the liquid. (Fig.~\ref{fig2}C). The fixed point was chosen at the midpoint of the sphere's trajectory in $y$ and 2.5 sphere radii to the right of the sphere's movement axis. The sphere's velocity profile was found as described in the Supplementary Material. As shown in Figure~\ref{fig2}C, there is a time delay $\tau$ between the sphere's velocity profile and the liquid's velocity profile, confirming the presence of inertial effects.

To track the position of the vortex over time, we again examine the velocity component in the direction of the sphere's movement ($v_y$). We first take $v_y$ along a line in $x$ that crosses through the center of the sphere, and each value of $v_y$ is averaged from 3 points in y. Then we take the vortex center to be the point where $v_y$ changes sign (Figure~\ref{fig2}E, see Supplementary Information for details of the calculations). This process is repeated for each frame of the video to track the location of the vortex over time. The appropriate first and last frames for this tracking are identified by visually comparing the calculated vortex center to the streamlines for each frame. The evolution of the vortex position in time is shown in Figure~\ref{fig2}F. 

For a Cyclet, the actuation process is very similar but with two Atwood and mass systems, as shown in Figure~\ref{fig3}A. The process used to kick the sphere upward is identical to the case of the single Pufflet. After the first mass lands, the string connected to the first mass is released, and the second mass is dropped. The second mass free-falls not under tension to accelerate quickly, then falls under tension for the last centimeter (approximately). While the second mass falls under tension, it applies a downward force to the sphere, kicking it back down to its starting location. Then, the mass is caught by a magnet, and the sphere comes to rest. 

This system allows the sphere to quickly accelerate upward, stop, then quickly accelerate downward, and finally stop at its starting location. This results in intermediate $\R_t$ numbers for the upward and downward motion of the sphere, while the $\R$ numbers always remain small. 

Individual tracer particles in the liquid were tracked using TrackPy\cite{allan_2025_16089574} (for Fig.~\ref{fig3}B) or TrackMate in ImageJ\cite{TrackMate7,TINEVEZ201780} (for Fig.~\ref{fig3}C). Any trajectories that did not persist through the entirety of the sphere's motion were discarded. The dataset shown in Figure~\ref{fig3}B was chosen for its clear and evenly distributed trajectories. 

Normalized displacements of particle trajectories were calculated by dividing each particle's final displacement by the distance it traveled after the first impulse (Fig.~\ref{fig3}C). The normalized displacement was calculated for all recorded trajectories from a single Cyclet for both Pufflets and Stokeslets. A different Cyclet dataset with a larger number of recorded trajectories was used in Figure~\ref{fig3}C for better statistics. 



\section*{Main Text Figures}

\begin{figure}[ht!]
 	\centering
\includegraphics[width=\textwidth]{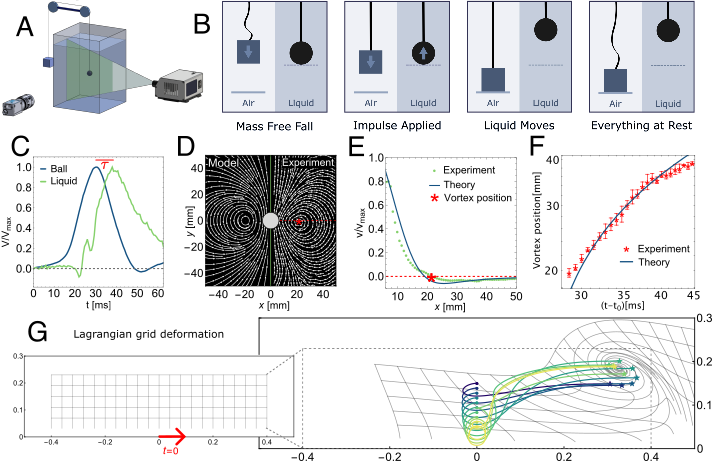}
 	\caption{\textbf{Measuring Pufflet flows}. 
    \textbf{(A)} Experimental set-up with Atwood machine, laser sheet, and high-speed camera.  
    \textbf{(B)} Cartoons describing the strong but short-lived force actuation. 
    \textbf{(C)} Velocity of the sphere as a function of time, compared to the velocity of the fluid at a point located at $x = 2.5 r$ and $y=z=0$. 
    \textbf{(D)} Comparison of experimental and theoretically modeled streamlines at $t=8$ ms (after the beginning of ball movement). 
    \textbf{(E)} Experimental and theoretical velocity profile along the red dashed line shown in panel D, at the same $t=8$ ms. 
    The vortex position is found as the root of the velocity profile. 
    \textbf{(F)} Experimental vortex positions against time, compared with theory. 
    \textbf{(G)} Simulation of Lagrangian grid displaced, distorted by a single Pufflet. Characteristic trajectories are plotted in color.}
    \label{fig2}
\end{figure}

\begin{figure}[ht!]
 	\centering
\includegraphics[width=0.8\textwidth]{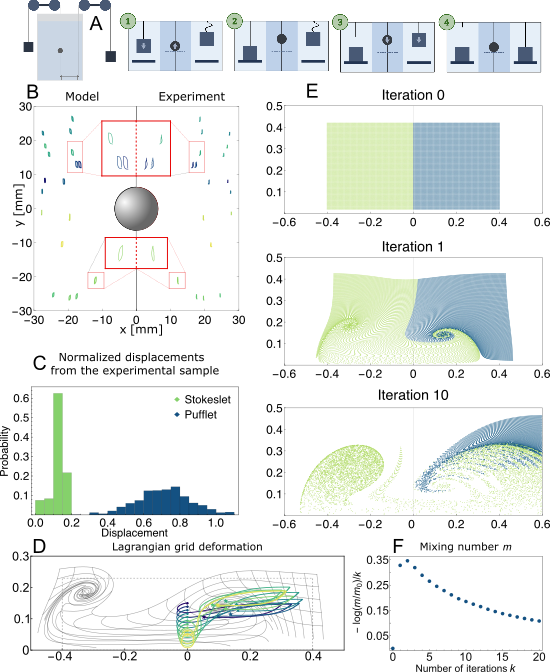}
\caption{\textbf{Irreversible mixing by cyclic forcing}. 
    \textbf{(A)} Experimental setup and schematic cartoon for a two-way forcing Cyclet experiment. 
    \textbf{(B)} Comparison of experimentally obtained and theoretically modeled trajectories starting at the same location. 
    \textbf{(C)} Histogram of normalized displacement after the whole cycle (two opposite impulses), for experimental Stokeslet and Pufflet cases. 
    \textbf{(D)} Lagrangian grid distortion and representative trajectories illustrating the time-irreversibility of Pufflets. 
    \textbf{(E)} Mixing by multiple iterations of Cyclets. 
    \textbf{(F)} Mixing number as a function of the number of iterations $k$, obtained from simulations presented in panel E.}
\label{fig3}
\end{figure}

\begin{figure}[ht!]
 	\centering
\includegraphics[width=\textwidth]{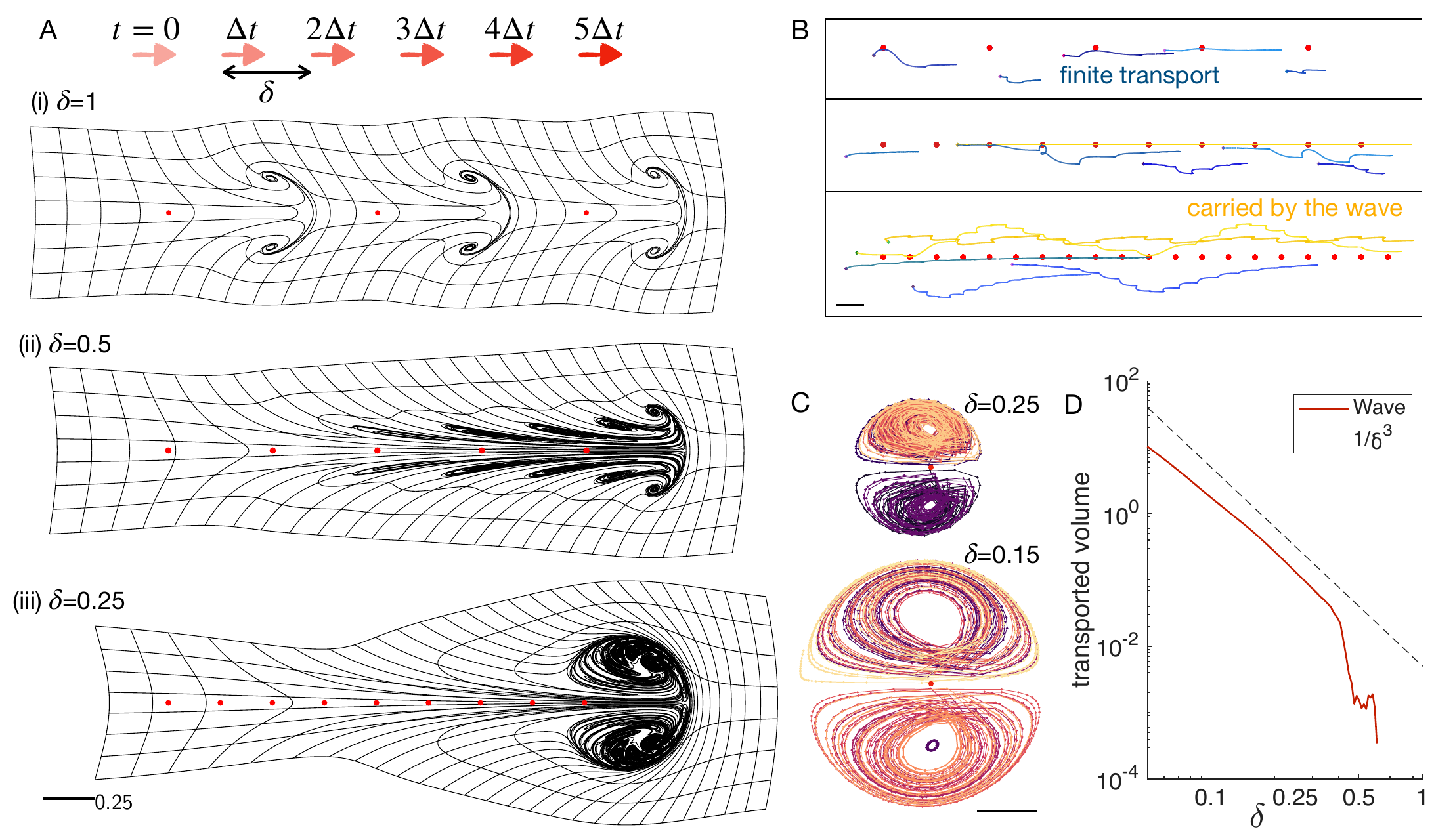}
 	\caption{ 
    \textbf{Long-ranged transport by surfing on a wave of Pufflets.}
    \textbf{(A)} Deformation of a square grid of fluid parcels by a wave of Pufflets with different spacings (see sketch), $\delta = 1, \, 0.5, \, \text{and} \, 0.25$. Scalebars in all panels are $0.25$. \textbf{(B)} Example trajectories for the same waves as in panel A. Yellow trajectories show transport along the waves, and blue ones display only transient motion. \textbf{(C)} Poincaré maps in the steady state for two different waves, plotted in the wave reference frame. Each trajectory shows the motion of a point transported by a wave, with the origin being the last active Pufflet. \textbf{(D)} Transported volume by a Pufflet wave, which is non-zero for sufficiently small spacing $\delta$ and scales as $\delta^{-3}$. }
    \label{fig4}
\end{figure}


\clearpage 

\bibliography{bib} 

@article{Bruot2016RealizingColloids,
    title = {{Realizing the Physics of Motile Cilia Synchronization with Driven Colloids}},
    year = {2016},
    journal = {Annual Review of Condensed Matter Physics},
    author = {Bruot, Nicolas and Cicuta, Pietro},
    number = {1},
    month = {3},
    pages = {323--348},
    volume = {7},
    publisher = {Annual Reviews Inc.},
    url = {https://www.annualreviews.org/doi/10.1146/annurev-conmatphys-031115-011451},
    doi = {10.1146/annurev-conmatphys-031115-011451},
    issn = {1947-5454}
}

@article{Lauga2009TheMicroorganisms,
    title = {{The hydrodynamics of swimming microorganisms}},
    year = {2009},
    journal = {Reports on Progress in Physics},
    author = {Lauga, Eric and Powers, Thomas R},
    number = {9},
    month = {9},
    pages = {096601},
    volume = {72},
    url = {https://iopscience.iop.org/article/10.1088/0034-4885/72/9/096601},
    doi = {10.1088/0034-4885/72/9/096601},
    issn = {0034-4885}
}

@article{Mathijssen2019,
  doi = {10.1038/s41586-019-1387-9},
  year = {2019},
  month = jul,
  publisher = {Springer Science and Business Media {LLC}},
  volume = {571},
  number = {7766},
  pages = {560--564},
  author = {Arnold J. T. M. Mathijssen and Joshua Culver and M. Saad Bhamla and Manu Prakash},
  title = {Collective intercellular communication through ultra-fast hydrodynamic trigger waves},
  journal = {Nature}
}

@article{Pozrikidis1989,
  doi = {10.1063/1.857329},
  year = {1989},
  month = sep,
  publisher = {{AIP} Publishing},
  volume = {1},
  number = {9},
  pages = {1508--1520},
  author = {C. Pozrikidis},
  title = {A singularity method for unsteady linearized flow},
  journal = {Physics of Fluids A: Fluid Dynamics}
}

@article{Fouxon_2018,
	author = {Fouxon, Itzhak and Leshansky, Alexander},
	doi = {10.1103/physreve.98.063108},
	issn = {2470-0053},
	journal = {Physical Review E},
	month = dec,
	number = {6},
	publisher = {American Physical Society (APS)},
	title = {Fundamental solution of unsteady Stokes equations and force on an oscillating sphere near a wall},
	url = {http://dx.doi.org/10.1103/PhysRevE.98.063108},
	volume = {98},
	year = {2018},
	}

@article{heimbichner_goebel_scaling_2020,
	title = {Scaling of ctenes and consequences for swimming performance in the ctenophore \textit{{Pleurobrachia} bachei}},
	volume = {139},
	issn = {1077-8306, 1744-7410},
	url = {https://onlinelibrary.wiley.com/doi/10.1111/ivb.12297},
	doi = {10.1111/ivb.12297},
	language = {en},
	number = {3},
	urldate = {2025-09-02},
	journal = {Invertebrate Biology},
	author = {Heimbichner Goebel, Wyatt L. and Colin, Sean P. and Costello, John H. and Gemmell, Brad J. and Sutherland, Kelly R.},
	month = sep,
	year = {2020},
	pages = {e12297},
}

@article{velho_rodrigues_bank_2021,
	title = {The bank of swimming organisms at the micron scale ({BOSO}-{Micro})},
	volume = {16},
	issn = {1932-6203},
	url = {https://dx.plos.org/10.1371/journal.pone.0252291},
	doi = {10.1371/journal.pone.0252291},
	language = {en},
	number = {6},
	urldate = {2025-09-02},
	journal = {PLOS ONE},
	author = {Velho Rodrigues, Marcos F. and Lisicki, Maciej and Lauga, Eric},
	editor = {Lele, Pushkar P},
	month = jun,
	year = {2021},
	pages = {e0252291},
}

@article{Aref1986,
  title = {Chaotic advection in a Stokes flow},
  volume = {29},
  ISSN = {0031-9171},
  url = {http://dx.doi.org/10.1063/1.865828},
  DOI = {10.1063/1.865828},
  number = {11},
  journal = {The Physics of Fluids},
  publisher = {AIP Publishing},
  author = {Aref,  H. and Balachandar,  S.},
  year = {1986},
  month = nov,
  pages = {3515–3521}
}

@article{biomimetics9040198,
    AUTHOR = {Cui, Zhiwei and Wang, Ye and den Toonder, Jaap M. J.},
    TITLE = {Metachronal Motion of Biological and Artificial Cilia},
    JOURNAL = {Biomimetics},
    VOLUME = {9},
    YEAR = {2024},
    NUMBER = {4},
    ARTICLE-NUMBER = {198},
    URL = {https://www.mdpi.com/2313-7673/9/4/198},
    PubMedID = {38667209},
    ISSN = {2313-7673},
    DOI = {10.3390/biomimetics9040198}
}

@article{ciliarelateddiseasesrev,
    author = {Ren, Zhanhong and Mao, Xiaoxiao and Wang, Siqi and Wang, Xin},
    title = {Cilia-related diseases},
    journal = {Journal of Cellular and Molecular Medicine},
    volume = {27},
    number = {24},
    pages = {3974-3979},
    doi = {https://doi.org/10.1111/jcmm.17990},
    url = {https://onlinelibrary.wiley.com/doi/abs/10.1111/jcmm.17990},
    eprint = {https://onlinelibrary.wiley.com/doi/pdf/10.1111/jcmm.17990},
    year = {2023}
}

@article{multiscalephys,
    author = {Gilpin, W. and Bull, M.S. and Prakash, M.},
    title = {The multiscale physics of cilia and flagella},
    journal = {Nature Reviews Physics},
    volume = {2},
    pages = {74–88},
    doi = {https://doi.org/10.1038/s42254-019-0129-0},
    url = {https://www.nature.com/articles/s42254-019-0129-0},
    year = {2020}
}

@article{Wei_Dehnavi_Aubin-Tam_Tam_2021, 
title={Measurements of the unsteady flow field around beating cilia}, 
volume={915}, 
DOI={10.1017/jfm.2021.149}, 
journal={Journal of Fluid Mechanics}, 
author={Wei, Da and Dehnavi, Parviz G. and Aubin-Tam, Marie-Eve and Tam, Daniel}, 
year={2021}, 
pages={A70}
}

@article{intReSquirmer,
    author = {Wang, S. and Ardekani, A.},
    title = {Biogenic mixing induced by intermediate Reynolds number swimming in stratified fluids},
    journal = {Scientific Reports},
    volume = {5},
    number = {17448},
    doi = {https://doi.org/10.1038/srep17448},
    url = {https://www.nature.com/articles/srep17448},
    year = {2015}
}

@article{TrackMate7,
    author = {Ershov, D. and Phan, MS. and Pylvänäinen, J.W. and Rigaud, S. U. and  Le Blanc, L. and Charles-Orszag, A. and Tinevez, J.-Y.},
    title = {TrackMate 7: integrating state-of-the-art segmentation algorithms into tracking pipelines},
    journal = {Nature Methods},
    volume = {19},
    pages = {829–832},
    doi = {https://doi.org/10.1038/s41592-022-01507-1},
    url = {https://www.nature.com/articles/s41592-022-01507-1},
    year = {2022}
}

@article{TINEVEZ201780,
title = {TrackMate: An open and extensible platform for single-particle tracking},
journal = {Methods},
volume = {115},
pages = {80-90},
year = {2017},
note = {Image Processing for Biologists},
issn = {1046-2023},
doi = {https://doi.org/10.1016/j.ymeth.2016.09.016},
url = {https://www.sciencedirect.com/science/article/pii/S1046202316303346},
author = {Jean-Yves Tinevez and Nick Perry and Johannes Schindelin and Genevieve M. Hoopes and Gregory D. Reynolds and Emmanuel Laplantine and Sebastian Y. Bednarek and Spencer L. Shorte and Kevin W. Eliceiri}
}

@article{Thielicke_2021, 
doi = {10.5334/jors.334}, 
url = {https://doi.org/10.5334%2Fjors.334}, 
year = 2021, 
month = {may}, 
publisher = {Ubiquity Press, Ltd.}, 
volume = {9}, 
number = {1}, 
pages = {12}, 
author = {William Thielicke and Ren{\'{e}} Sonntag}, 
title = {Particle Image Velocimetry for {MATLAB}: Accuracy and enhanced algorithms in {PIVlab}}, 
journal = {Journal of Open Research Software}
}

@article{PhysRevLett.96.058102,
  title = {Hydrodynamic Flow Patterns and Synchronization of Beating Cilia},
  author = {Vilfan, Andrej and J\"ulicher, Frank},
  journal = {Phys. Rev. Lett.},
  volume = {96},
  issue = {5},
  pages = {058102},
  numpages = {4},
  year = {2006},
  month = {Feb},
  publisher = {American Physical Society},
  doi = {10.1103/PhysRevLett.96.058102},
  url = {https://link.aps.org/doi/10.1103/PhysRevLett.96.058102}
}

@article{10.1126/science.aae0450,
author = {Faubel, Regina and Westendorf, Christian and Bodenschatz, Eberhard and Eichele, Gregor},
title = {Cilia-based flow network in the brain ventricles},
journal = {Science},
volume = {353},
number = {6295},
pages = {176-178},
year = {2016},
doi = {10.1126/science.aae0450},
URL = {https://www.science.org/doi/abs/10.1126/science.aae0450},
eprint = {https://www.science.org/doi/pdf/10.1126/science.aae0450}
}

@article{SpontaneousCreation,
author = {Guirao, Boris and Joanny, Jean-François},
title = {Spontaneous Creation of Macroscopic Flow and Metachronal Waves in an Array of Cilia},
journal = {Biophysical Journal},
volume = {92},
number = {6},
pages = {1900-1917},
year = {2007},
doi = {10.1529/biophysj.106.084897},
URL = {https://www.cell.com/AJHG/fulltext/S0006-3495%2807%2970998-0}
}

@article{doi:10.1126/sciadv.abd2508,
author = {Edoardo Milana  and Rongjing Zhang  and Maria Rosaria Vetrano  and Sam Peerlinck  and Michael De Volder  and Patrick R. Onck  and Dominiek Reynaerts  and Benjamin Gorissen },
title = {Metachronal patterns in artificial cilia for low Reynolds number fluid propulsion},
journal = {Science Advances},
volume = {6},
number = {49},
pages = {eabd2508},
year = {2020},
doi = {10.1126/sciadv.abd2508},
URL = {https://www.science.org/doi/abs/10.1126/sciadv.abd2508},
eprint = {https://www.science.org/doi/pdf/10.1126/sciadv.abd2508}
}

@article{10.1038/s41598-020-64420-7,
title = {Ciliary vortex flows and oxygen dynamics in the coral boundary layer},
journal = {Scientific Reports},
volume = {10},
number = {7541},
year = {2020},
doi = {10.1038/s41598-020-64420-7},
url = {https://www.nature.com/articles/s41598-020-64420-7},
author = {Cesar O. Pacherres and Soeren Ahmerkamp and Gertraud M. Schmidt-Grieb and Moritz Holtappels and Claudio Richter} 
}

@article{doi:10.1073/pnas.1218869110,
author = {Jens Elgeti  and Gerhard Gompper},
title = {Emergence of metachronal waves in cilia arrays},
journal = {Proceedings of the National Academy of Sciences},
volume = {110},
number = {12},
pages = {4470-4475},
year = {2013},
doi = {10.1073/pnas.1218869110},
URL = {https://www.pnas.org/doi/abs/10.1073/pnas.1218869110},
eprint = {https://www.pnas.org/doi/pdf/10.1073/pnas.1218869110}
}

@article{doi:10.1073/pnas.2214413119,
author = {Anup V. Kanale  and Feng Ling  and Hanliang Guo  and Sebastian Fürthauer  and Eva Kanso},
title = {Spontaneous phase coordination and fluid pumping in model ciliary carpets},
journal = {Proceedings of the National Academy of Sciences},
volume = {119},
number = {45},
pages = {e2214413119},
year = {2022},
doi = {10.1073/pnas.2214413119},
URL = {https://www.pnas.org/doi/abs/10.1073/pnas.2214413119},
eprint = {https://www.pnas.org/doi/pdf/10.1073/pnas.2214413119}
}

@article{Byron2021MetachronalDirections,
    title = {{Metachronal Motion across Scales: Current Challenges and Future Directions}},
    year = {2021},
    journal = {Integrative and Comparative Biology},
    author = {Byron, Margaret L and Murphy, David W and Katija, Kakani and Hoover, Alexander P and Daniels, Joost and Garayev, Kuvvat and Takagi, Daisuke and Kanso, Eva and Gemmell, Bradford J and Ruszczyk, Melissa and Santhanakrishnan, Arvind},
    number = {5},
    month = {11},
    pages = {1674--1688},
    volume = {61},
    url = {https://academic.oup.com/icb/article/61/5/1674/6287617},
    doi = {10.1093/icb/icab105},
    issn = {1540-7063}
}

@article{Nawroth2020MultiscaleLung,
    title = {{Multiscale mechanics of mucociliary clearance in the lung}},
    year = {2020},
    journal = {Philosophical Transactions of the Royal Society B: Biological Sciences},
    author = {Nawroth, Janna C. and van der Does, Anne M. and Ryan (Firth), Amy and Kanso, Eva},
    number = {1792},
    month = {2},
    pages = {20190160},
    volume = {375},
    url = {https://royalsocietypublishing.org/doi/10.1098/rstb.2019.0160},
    doi = {10.1098/rstb.2019.0160},
    issn = {0962-8436}
}

@article{Ramirez-SanJuan2020Multi-scaleArrays,
    title = {{Multi-scale spatial heterogeneity enhances particle clearance in airway ciliary arrays}},
    year = {2020},
    journal = {Nature Physics},
    author = {Ramirez-San Juan, Guillermina R. and Mathijssen, Arnold J. T. M. and He, Mu and Jan, Lily and Marshall, Wallace and Prakash, Manu},
    number = {9},
    month = {9},
    pages = {958--964},
    volume = {16},
    publisher = {Nature Research},
    url = {https://www.nature.com/articles/s41567-020-0923-8},
    doi = {10.1038/s41567-020-0923-8},
    issn = {1745-2473}
}

@article{Worthington1963,
  title = {Ependymal Cilia: Distribution and Activity in the Adult Human Brain},
  volume = {139},
  ISSN = {1095-9203},
  url = {http://dx.doi.org/10.1126/science.139.3551.221},
  DOI = {10.1126/science.139.3551.221},
  number = {3551},
  journal = {Science},
  publisher = {American Association for the Advancement of Science (AAAS)},
  author = {Worthington,  W. Curtis and Cathcart,  Robert S.},
  year = {1963},
  month = jan,
  pages = {221–222}
}

@article{Solovev_2022,
doi = {10.1088/1367-2630/ac2ae4},
url = {https://doi.org/10.1088/1367-2630/ac2ae4},
year = {2022},
month = {jan},
publisher = {IOP Publishing},
volume = {24},
number = {1},
pages = {013015},
author = {Solovev, Anton and Friedrich, Benjamin M},
title = {Synchronization in cilia carpets: multiple metachronal waves are stable, but one wave dominates},
journal = {New Journal of Physics}
}

@article{B901660K,
author ={Baltussen, Michiel and Anderson, Patrick and Bos, Femke and den Toonder, Jaap},
title  ={Inertial flow effects in a micro-mixer based on artificial cilia},
journal  ={Lab Chip},
year  ={2009},
volume  ={9},
issue  ={16},
pages  ={2326-2331},
publisher  ={The Royal Society of Chemistry},
doi  ={10.1039/B901660K},
url  ={http://dx.doi.org/10.1039/B901660K}
}

@article{PhysRevLett.122.124502,
  title = {Is the Zero Reynolds Number Approximation Valid for Ciliary Flows?},
  author = {Wei, Da and Dehnavi, Parviz Ghoddoosi and Aubin-Tam, Marie-Eve and Tam, Daniel},
  journal = {Phys. Rev. Lett.},
  volume = {122},
  issue = {12},
  pages = {124502},
  numpages = {5},
  year = {2019},
  month = {Mar},
  publisher = {American Physical Society},
  doi = {10.1103/PhysRevLett.122.124502},
  url = {https://link.aps.org/doi/10.1103/PhysRevLett.122.124502}
}

@article{PhysRevResearch.7.L012029,
  title = {Synchronization and metachronal waves of elastic cilia caused by unsteady viscous flow},
  author = {von Kenne, Albert and Schmelter, Sonja and Stark, Holger and B\"ar, Markus},
  journal = {Phys. Rev. Res.},
  volume = {7},
  issue = {1},
  pages = {L012029},
  numpages = {8},
  year = {2025},
  month = {Feb},
  publisher = {American Physical Society},
  doi = {10.1103/PhysRevResearch.7.L012029},
  url = {https://link.aps.org/doi/10.1103/PhysRevResearch.7.L012029}
}

@incollection{10.1115/IMECE2023-112572,
    author = {Lei, Menglong and Lou, Zhipeng and Wang, Junshi and Dong, Haibo and Li, Chengyu},
    title = {Hydrodynamics of Metachronal Rowing at Intermediate Reynolds Numbers},
    volume = {Volume 9: Fluids Engineering},
    series = {ASME International Mechanical Engineering Congress and Exposition},
    pages = {V009T10A006},
    year = {2023},
    month = {10},
    doi = {10.1115/IMECE2023-112572},
    url = {https://doi.org/10.1115/IMECE2023-112572},
    eprint = {https://asmedigitalcollection.asme.org/IMECE/proceedings-pdf/IMECE2023/87660/V009T10A006/7239756/v009t10a006-imece2023-112572.pdf},
}

@article{PhysRevFluids.5.063103,
  title = {Kinematics of a simple reciprocal model swimmer at intermediate Reynolds numbers},
  author = {Dombrowski, Thomas and Klotsa, Daphne},
  journal = {Phys. Rev. Fluids},
  volume = {5},
  issue = {6},
  pages = {063103},
  numpages = {20},
  year = {2020},
  month = {Jun},
  publisher = {American Physical Society},
  doi = {10.1103/PhysRevFluids.5.063103},
  url = {https://link.aps.org/doi/10.1103/PhysRevFluids.5.063103}
}

@article{10.1093/icb/icab179,
    title = {{Spatiotemporal Asymmetry in Metachronal Rowing at Intermediate Reynolds Numbers}},
    year = {2021},
    journal = {Integrative and Comparative Biology},
    author = {Herrera-Amaya, Adrian and Seber, Elizabeth K and Murphy, David W and Patry, Wyatt L and Knowles, Thomas S and Bubel, MacKenzie M and Maas, Amy E and Byron, Margaret L},
    number = {5},
    month = {11},
    pages = {1579--1593},
    volume = {61},
    url = {https://academic.oup.com/icb/article/61/5/1579/6354788},
    doi = {10.1093/icb/icab179},
    issn = {1540-7063}
}

@article{PhysRevFluids.6.053102,
  title = {Direct measurement of unsteady microscale Stokes flow using optically driven microspheres},
  author = {Bruot, Nicolas and Cicuta, Pietro and Bloomfield-Gad\^elha, Hermes and Goldstein, Raymond E. and Kotar, Jurij and Lauga, Eric and Nadal, Fran\ifmmode \mbox{\c{c}}\else \c{c}\fi{}ois},
  journal = {Phys. Rev. Fluids},
  volume = {6},
  issue = {5},
  pages = {053102},
  numpages = {14},
  year = {2021},
  month = {May},
  publisher = {American Physical Society},
  doi = {10.1103/PhysRevFluids.6.053102},
  url = {https://link.aps.org/doi/10.1103/PhysRevFluids.6.053102}
}

@article{Herrera-Amaya_2024,
doi = {10.1088/1748-3190/ad7abf},
url = {https://doi.org/10.1088/1748-3190/ad7abf},
year = {2024},
month = {oct},
publisher = {IOP Publishing},
volume = {19},
number = {6},
pages = {066004},
author = {Herrera-Amaya, Adrian and Byron, Margaret L},
title = {Propulsive efficiency of spatiotemporally asymmetric oscillating appendages at intermediate Reynolds numbers},
journal = {Bioinspiration \& Biomimetics}
}

@article{C4SM00770K,
author ={Theers, Mario and Winkler, Roland G.},
title  ={Effects of thermal fluctuations and fluid compressibility on hydrodynamic synchronization of microrotors at finite oscillatory Reynolds number: a multiparticle collision dynamics simulation study},
journal  ={Soft Matter},
year  ={2014},
volume  ={10},
issue  ={32},
pages  ={5894-5904},
publisher  ={The Royal Society of Chemistry},
doi  ={10.1039/C4SM00770K},
url  ={http://dx.doi.org/10.1039/C4SM00770K}
}

@book{pozrikidis_introduction_1997,
	title = {Introduction to Theoretical and Computational Fluid Dynamics},
	language = {en},
    publisher = {Oxford University Press},
	author = {Pozrikidis, Costas and Ferziger, Joel H.},
	month = sep,
	year = {2011},
    ISBN = {9780199752072}
}

@article{DiCarlo2009InertialMicrofluidics,
    title = {{Inertial microfluidics}},
    year = {2009},
    journal = {Lab on a Chip},
    author = {Di Carlo, Dino},
    number = {21},
    pages = {3038},
    volume = {9},
    url = {https://xlink.rsc.org/?DOI=b912547g},
    doi = {10.1039/b912547g},
    issn = {1473-0197}
}

@article{Zhang2016FundamentalsReview,
    title = {{Fundamentals and applications of inertial microfluidics: a review}},
    year = {2016},
    journal = {Lab on a Chip},
    author = {Zhang, Jun and Yan, Sheng and Yuan, Dan and Alici, Gursel and Nguyen, Nam-Trung and Ebrahimi Warkiani, Majid and Li, Weihua},
    number = {1},
    pages = {10--34},
    volume = {16},
    url = {https://xlink.rsc.org/?DOI=C5LC01159K},
    doi = {10.1039/C5LC01159K},
    issn = {1473-0197}
}

@article{Peterman2024EncodingPlatform,
    title = {{Encoding spatiotemporal asymmetry in artificial cilia with a ctenophore-inspired soft-robotic platform}},
    year = {2024},
    journal = {Bioinspiration {\&} Biomimetics},
    author = {Peterman, David J and Byron, Margaret L},
    number = {6},
    month = {11},
    pages = {066002},
    volume = {19},
    url = {https://iopscience.iop.org/article/10.1088/1748-3190/ad791c},
    doi = {10.1088/1748-3190/ad791c},
    issn = {1748-3182}
}

@article{Demirors2021AmphibiousCarpets,
    title = {{Amphibious Transport of Fluids and Solids by Soft Magnetic Carpets}},
    year = {2021},
    journal = {Advanced Science},
    author = {Demir{\"{o}}rs, Ahmet F. and Aykut, Sümeyye and Ganzeboom, Sophia and Meier, Yuki A. and Hardeman, Robert and de Graaf, Joost and Mathijssen, Arnold J. T. M. and Poloni, Erik and Carpenter, Julia A. and {\"{U}}nl{\"{u}}, Caner and Zenh{\"{a}}usern, Daniel},
    number = {21},
    month = {11},
    pages = {2102510},
    volume = {8},
    url = {https://advanced.onlinelibrary.wiley.com/doi/10.1002/advs.202102510},
    doi = {10.1002/advs.202102510},
    issn = {2198-3844}
}

@article{Ford2019HydrodynamicsLag,
    title = {{Hydrodynamics of metachronal paddling: effects of varying Reynolds number and phase lag}},
    year = {2019},
    journal = {Royal Society Open Science},
    author = {Ford, Mitchell P. and Lai, Hong Kuan and Samaee, Milad and Santhanakrishnan, Arvind},
    number = {10},
    month = {10},
    pages = {191387},
    volume = {6},
    url = {https://royalsocietypublishing.org/doi/10.1098/rsos.191387},
    doi = {10.1098/rsos.191387},
    issn = {2054-5703}
}

@book{taylor1967video,
	title        = {Low-{R}eynolds-number flows},
	author       = {Taylor, G. I.},
	year         = 1967,
	publisher    = {National Committee for Fluid Mechanics Films},
	url          = {http://web.mit.edu/hml/ncfmf.html},
	note         = {Last accessed: 16-Dec-2020}
}

@article{Brennen1977FluidFlagella,
    title = {{Fluid Mechanics of Propulsion by Cilia and Flagella}},
    year = {1977},
    journal = {Annual Review of Fluid Mechanics},
    author = {Brennen, C and Winet, H},
    number = {1},
    month = {1},
    pages = {339--398},
    volume = {9},
    publisher = {Annual Reviews},
    url = {https://www.annualreviews.org/doi/10.1146/annurev.fl.09.010177.002011},
    doi = {10.1146/annurev.fl.09.010177.002011},
    issn = {0066-4189}
}

@article{ulIslam2022MicroscopicReview,
    title = {{Microscopic artificial cilia – a review}},
    year = {2022},
    journal = {Lab on a Chip},
    author = {ul Islam, Tanveer and Wang, Ye and Aggarwal, Ishu and Cui, Zhiwei and Eslami Amirabadi, Hossein and Garg, Hemanshul and Kooi, Roel and Venkataramanachar, Bhavana B. and Wang, Tongsheng and Zhang, Shuaizhong and Onck, Patrick R. and den Toonder, Jaap M. J.},
    number = {9},
    pages = {1650--1679},
    volume = {22},
    url = {https://xlink.rsc.org/?DOI=D1LC01168E},
    doi = {10.1039/D1LC01168E},
    issn = {1473-0197}
}

@article{Toonder2008ArtificialMixing,
    title = {{Artificial cilia for active micro-fluidic mixing}},
    year = {2008},
    journal = {Lab on a Chip},
    author = {Toonder, Jaap den and Bos, Femke and Broer, Dick and Filippini, Laura and Gillies, Murray and de Goede, Judith and Mol, Titie and Reijme, Mireille and Talen, Wim and Wilderbeek, Hans and Khatavkar, Vinayak and Anderson, Patrick},
    number = {4},
    month = {3},
    pages = {533},
    volume = {8},
    publisher = {Royal Society of Chemistry},
    url = {https://xlink.rsc.org/?DOI=b717681c},
    doi = {10.1039/b717681c},
    issn = {1473-0197}
}

@article{Winkler2016LowSimulations,
    title = {{Low Reynolds number hydrodynamics and mesoscale simulations}},
    year = {2016},
    journal = {Eur. Phys. J. Spec. Top.},
    author = {Winkler, R. G.},
    pages = {2079--2097},
    volume = {225},
    doi={10.1140/epjst/e2016-60087-9},
url={http://link.springer.com/10.1140/epjst/e2016-60087-9}
}

@article{Ilton2018TheSystems,
    title = {{The principles of cascading power limits in small, fast biological and engineered systems}},
    year = {2018},
    journal = {Science},
    author = {Ilton, Mark and Bhamla, M. Saad and Ma, Xiaotian and Cox, Suzanne M. and Fitchett, Leah L. and Kim, Yongjin and Koh, Je-sung and Krishnamurthy, Deepak and Kuo, Chi-Yun and Temel, Fatma Zeynep and Crosby, Alfred J. and Prakash, Manu and Sutton, Gregory P. and Wood, Robert J. and Azizi, Emanuel and Bergbreiter, Sarah and Patek, S. N.},
    number = {6387},
    month = {4},
    volume = {360},
    url = {https://www.science.org/doi/10.1126/science.aao1082},
    doi = {10.1126/science.aao1082},
    issn = {0036-8075}
}

@article{Hubert2021ScallopMesoscale,
    title = {{Scallop Theorem and Swimming at the Mesoscale}},
    year = {2021},
    journal = {Physical Review Letters},
    author = {Hubert, M. and Trosman, O. and Collard, Y. and Sukhov, A. and Harting, J. and Vandewalle, N. and Smith, A.-S.},
    number = {22},
    month = {6},
    pages = {224501},
    volume = {126},
    url = {https://link.aps.org/doi/10.1103/PhysRevLett.126.224501},
    doi = {10.1103/PhysRevLett.126.224501},
    issn = {0031-9007}
}

@misc{allan_2025_16089574,
  author    = {Allan, Daniel B. and Caswell, Thomas and Keim, Nathan C. and van der Wel, Casper M. and Verweij, Ruben W.},
  title     = {{soft-matter/trackpy}: v0.7},
  year      = {2025},
  month     = jul,
  publisher = {Zenodo},
  version   = {v0.7},
  doi       = {10.5281/zenodo.16089574},
  url       = {https://doi.org/10.5281/zenodo.16089574}
}

@article{Nagai2013,
  title = {Mixing of solutions by coordinated ciliary motion in Vorticella convallaria and patterning method for microfluidic applications},
  volume = {188},
  ISSN = {0925-4005},
  url = {http://dx.doi.org/10.1016/j.snb.2013.08.040},
  DOI = {10.1016/j.snb.2013.08.040},
  journal = {Sensors and Actuators B: Chemical},
  publisher = {Elsevier BV},
  author = {Nagai,  Moeto and Hayasaka,  Yo and Kato,  Kei and Kawashima,  Takahiro and Shibata,  Takayuki},
  year = {2013},
  month = nov,
  pages = {1255–1262}
}

@article{Sears2013,
  title = {Human airway ciliary dynamics},
  volume = {304},
  ISSN = {1522-1504},
  url = {http://dx.doi.org/10.1152/ajplung.00105.2012},
  DOI = {10.1152/ajplung.00105.2012},
  number = {3},
  journal = {American Journal of Physiology-Lung Cellular and Molecular Physiology},
  publisher = {American Physiological Society},
  author = {Sears,  Patrick R. and Thompson,  Kristin and Knowles,  Michael R. and Davis,  C. William},
  year = {2013},
  month = feb,
  pages = {L170–L183}
}

@article{Lyons2006,
  title = {The reproductive significance of human Fallopian tube cilia},
  volume = {12},
  ISSN = {1355-4786},
  url = {http://dx.doi.org/10.1093/humupd/dml012},
  DOI = {10.1093/humupd/dml012},
  number = {4},
  journal = {Human Reproduction Update},
  publisher = {Oxford University Press (OUP)},
  author = {Lyons,  R.A. and Saridogan,  E. and Djahanbakhch,  O.},
  year = {2006},
  month = mar,
  pages = {363–372}
}

@article{Zhou2017,
  title = {Flow and transport effect caused by the stalk contraction cycle of Vorticella convallaria},
  volume = {11},
  ISSN = {1932-1058},
  url = {http://dx.doi.org/10.1063/1.4985654},
  DOI = {10.1063/1.4985654},
  number = {3},
  journal = {Biomicrofluidics},
  publisher = {AIP Publishing},
  author = {Zhou,  Jiazhong and Ryu,  Sangjin and Admiraal,  David},
  year = {2017},
  month = may 
}

@article{Ryu2010,
  title = {Unsteady Motion,  Finite Reynolds Numbers,  and Wall Effect on Vorticella convallaria Contribute Contraction Force Greater than the Stokes Drag},
  volume = {98},
  ISSN = {0006-3495},
  url = {http://dx.doi.org/10.1016/j.bpj.2010.02.025},
  DOI = {10.1016/j.bpj.2010.02.025},
  number = {11},
  journal = {Biophysical Journal},
  publisher = {Elsevier BV},
  author = {Ryu,  Sangjin and Matsudaira,  Paul},
  year = {2010},
  month = jun,
  pages = {2574–2581}
}

@article{ding_mixing_2014,
	title = {Mixing and transport by ciliary carpets: a numerical study},
	volume = {743},
	copyright = {https://www.cambridge.org/core/terms},
	issn = {0022-1120, 1469-7645},
	shorttitle = {Mixing and transport by ciliary carpets},
	url = {https://www.cambridge.org/core/product/identifier/S0022112014000366/type/journal_article},
	doi = {10.1017/jfm.2014.36},
	language = {en},
	urldate = {2025-08-04},
	journal = {Journal of Fluid Mechanics},
	author = {Ding, Yang and Nawroth, Janna C. and McFall-Ngai, Margaret J. and Kanso, Eva},
	month = mar,
	year = {2014},
	pages = {124--140},
}

@article{krasnopolskaya_mixing_1999,
	title = {Mixing in {Stokes} flow in an annular wedge cavity},
	volume = {18},
	copyright = {https://www.elsevier.com/tdm/userlicense/1.0/},
	issn = {09977546},
	url = {https://linkinghub.elsevier.com/retrieve/pii/S0997754699001193},
	doi = {10.1016/S0997-7546(99)00119-3},
	language = {en},
	number = {5},
	urldate = {2025-12-09},
	journal = {European Journal of Mechanics - B/Fluids},
	author = {Krasnopolskaya, T.S. and Meleshko, V.V. and Peters, G.W.M. and Meijer, H.E.H.},
	month = sep,
	year = {1999},
	pages = {793--822},
}

@article{stone_imaging_2005,
	title = {Imaging and quantifying mixing in a model droplet micromixer},
	volume = {17},
	issn = {1070-6631, 1089-7666},
	url = {https://pubs.aip.org/pof/article/17/6/063103/256055/Imaging-and-quantifying-mixing-in-a-model-droplet},
	doi = {10.1063/1.1929547},
	language = {en},
	number = {6},
	urldate = {2025-12-09},
	journal = {Physics of Fluids},
	author = {Stone, Z. B. and Stone, H. A.},
	month = jun,
	year = {2005},
	pages = {063103},
}

@article{mathew_multiscale_2005,
	title = {A multiscale measure for mixing},
	volume = {211},
	copyright = {https://www.elsevier.com/tdm/userlicense/1.0/},
	issn = {01672789},
	url = {https://linkinghub.elsevier.com/retrieve/pii/S0167278905003313},
	doi = {10.1016/j.physd.2005.07.017},
	language = {en},
	number = {1-2},
	urldate = {2025-12-09},
	journal = {Physica D: Nonlinear Phenomena},
	author = {Mathew, George and Mezić, Igor and Petzold, Linda},
	month = nov,
	year = {2005},
	pages = {23--46},
}

@article{reece_new_2020,
	title = {New instability and mixing simulations using {SPH} and a novel mixing measure},
	volume = {32},
	issn = {1001-6058, 1878-0342},
	url = {https://link.springer.com/10.1007/s42241-020-0045-x},
	doi = {10.1007/s42241-020-0045-x},
	language = {en},
	number = {4},
	urldate = {2025-12-09},
	journal = {Journal of Hydrodynamics},
	author = {Reece, Georgina and Rogers, Benedict D. and Lind, Steven and Fourtakas, Georgios},
	month = aug,
	year = {2020},
	pages = {684--698},
}

@phdthesis{lyapunov,
  author       = {Lyapunov, Aleksandr Mikhailovich},
  title        = {The General Problem of the Stability of Motion},
  school       = {University of Kharkov},
  address      = {Kharkov},
  year         = {1892},
  pages        = {250},
  language     = {Russian},
  note         = {Doctoral dissertation, Kharkov Mathematical Society}
}

@article{https://doi.org/10.1002/adfm.201706666,
author = {Ben, Shuang and Tai, Jun and Ma, Han and Peng, Yun and Zhang, Yuan and Tian, Dongliang and Liu, Kesong and Jiang, Lei},
title = {Cilia-Inspired Flexible Arrays for Intelligent Transport of Viscoelastic Microspheres},
journal = {Advanced Functional Materials},
volume = {28},
number = {16},
pages = {1706666},
doi = {https://doi.org/10.1002/adfm.201706666},
url = {https://advanced.onlinelibrary.wiley.com/doi/abs/10.1002/adfm.201706666},
eprint = {https://advanced.onlinelibrary.wiley.com/doi/pdf/10.1002/adfm.201706666},
year = {2018}
}

@article{benBioinspiredMagneticResponsive2020,
	title = {A bioinspired magnetic responsive cilia array surface for microspheres underwater directional transport},
	volume = {63},
	issn = {1869-1870},
	url = {https://doi.org/10.1007/s11426-019-9660-5},
	doi = {10.1007/s11426-019-9660-5},
	number = {3},
	journal = {Science China Chemistry},
	author = {Ben, Shuang and Yao, Jinjia and Ning, Yuzhen and Zhao, Zhihong and Zha, Jinlong and Tian, Dongliang and Liu, Kesong and Jiang, Lei},
	month = mar,
	year = {2020},
	pages = {347--353},
}

@article{doi:10.1021/acsami.0c10034,
author = {Song, Yuegan and Jiang, Shaojun and Li, Guoqiang and Zhang, Yachao and Wu, Hao and Xue, Cheng and You, Hongshu and Zhang, Dehu and Cai, Yong and Zhu, Jiangong and Zhu, Wulin and Li, Jiawen and Hu, Yanlei and Wu, Dong and Chu, Jiaru},
title = {Cross-Species Bioinspired Anisotropic Surfaces for Active Droplet Transportation Driven by Unidirectional Microcolumn Waves},
journal = {ACS Applied Materials \& Interfaces},
volume = {12},
number = {37},
pages = {42264-42273},
year = {2020},
doi = {10.1021/acsami.0c10034},
URL = {https://doi.org/10.1021/acsami.0c10034},
eprint = {https://doi.org/10.1021/acsami.0c10034}
}

@article{doi:10.1021/acsami.1c03009,
author = {Zhang, Shuaizhong and Cui, Zhiwei and Wang, Ye and den Toonder, Jaap},
title = {Metachronal $\mu$-Cilia for On-Chip Integrated Pumps and Climbing Robots},
journal = {ACS Applied Materials \& Interfaces},
volume = {13},
number = {17},
pages = {20845-20857},
year = {2021},
doi = {10.1021/acsami.1c03009},
URL = {https://doi.org/10.1021/acsami.1c03009},
eprint = {https://doi.org/10.1021/acsami.1c03009}
}

@article{guMagneticCiliaCarpets2020a,
  title = {Magnetic Cilia Carpets with Programmable Metachronal Waves},
  author = {Gu, Hongri and Boehler, Quentin and Cui, Haoyang and Secchi, Eleonora and Savorana, Giovanni and De Marco, Carmela and Gervasoni, Simone and Peyron, Quentin and Huang, Tian-Yun and Pane, Salvador and Hirt, Ann M. and Ahmed, Daniel and Nelson, Bradley J.},
  year = {2020},
  month = may,
  journal = {Nature Communications},
  volume = {11},
  number = {1},
  pages = {2637},
  issn = {2041-1723},
  doi = {10.1038/s41467-020-16458-4},
}

@Article{D1SM01680F,
author ={Zhang, Rongjing and Toonder, Jaap den and Onck, Patrick R.},
title  ={Metachronal patterns by magnetically-programmable artificial cilia surfaces for low Reynolds number fluid transport and mixing},
journal  ={Soft Matter},
year  ={2022},
volume  ={18},
issue  ={20},
pages  ={3902-3909},
publisher  ={The Royal Society of Chemistry},
doi  ={10.1039/D1SM01680F},
url  ={http://dx.doi.org/10.1039/D1SM01680F},
}
\bibliographystyle{sciencemag}


\newpage


\renewcommand{\thefigure}{S\arabic{figure}}
\renewcommand{\thetable}{S\arabic{table}}
\renewcommand{\theequation}{S\arabic{equation}}
\renewcommand{\thepage}{S\arabic{page}}
\setcounter{figure}{0}
\setcounter{table}{0}
\setcounter{equation}{0}
\setcounter{page}{1} 


\begin{center}
\section*{Supplementary Materials for\\ \scititle}


\author{
    Rafał Błaszkiewicz$^{1\dagger}$,
    Margot Young$^{2\dagger}$,
    Albane Th{\'e}ry$^{2,3\ast}$, 
    Talia Becker Calazans$^{2}$,
    Yoichiro Mori$^{3}$,
    Maciej Lisicki$^{1\ast}$, 
	Arnold J. T. M. Mathijssen$^{2\ast}$\\
	{\small{$^\ast$Corresponding authors. Emails: albane.thery@warwick.ac.uk (AT),  mklis@fuw.edu.pl (ML) and amaths@upenn.edu (AJTMM).}}\\
	\small$^\dagger$These authors contributed equally to this work.
}
\end{center}

\subsubsection*{Experimental $\R$ \& $\R_t$ Analysis} \label{ReRetAnalysis}
For our $\R$ and $\R_t$ analysis, we use $\R$ and $\R_t$ as defined in Eq.~\eqref{ReDefs} , where $L_0$ is the sphere's diameter, $\nu$ is the silicone oil viscosity, $U_0$ is the sphere's characteristic velocity, and $T_{acc}$ is the time scale of the sphere's acceleration. The characteristic quantities $U_0$ and $T_{acc}$ are found from the motion of the sphere. 

We track the centroid of the sphere in each video using ImageJ's Hough Circle Transform. These trajectories are smoothed using a Gaussian kernel and used to calculate velocity and acceleration curves. 

For the velocity curves, we take $U_0 = U_{max}$ and $T_{acc}$ to be the time between the sphere's half-max and max velocities, as shown in Figure~\ref{fig:Re_Calcs}A. For the acceleration curves (Fig.~\ref{fig:Re_Calcs}B), we fit the experimental data to a double-Gaussian of the form $$f(t) = \frac{N_1}{\sigma_1 \sqrt{2\pi}}e^{-\frac{(t - \mu_1)^2}{2\sigma_1^2}} - \frac{N_1}{\sigma_2\sqrt{2\pi}}e^{-\frac{(t - \mu_2)^2}{2\sigma^2}}.$$ 
Using this fit, we take $$U_0=\int_{-\infty}^{x_{intercept}}f(t)dt$$ and $T_{acc}$ to be the time between the max acceleration and $a=0$. The final reported $\R$ and $\R_t$ values in the main text are the averages of the values from the velocity and acceleration curves. 

The sphere tracking and $\R$, $\R_t$ analysis for the Cyclets is identical to the single Pufflet case. The sphere tracking for the Cyclets was additionally used to identify when the sphere returns to its starting location, and videos were trimmed accordingly. 

\begin{figure}[h!]
    \centering
    \includegraphics[width=0.5\linewidth]{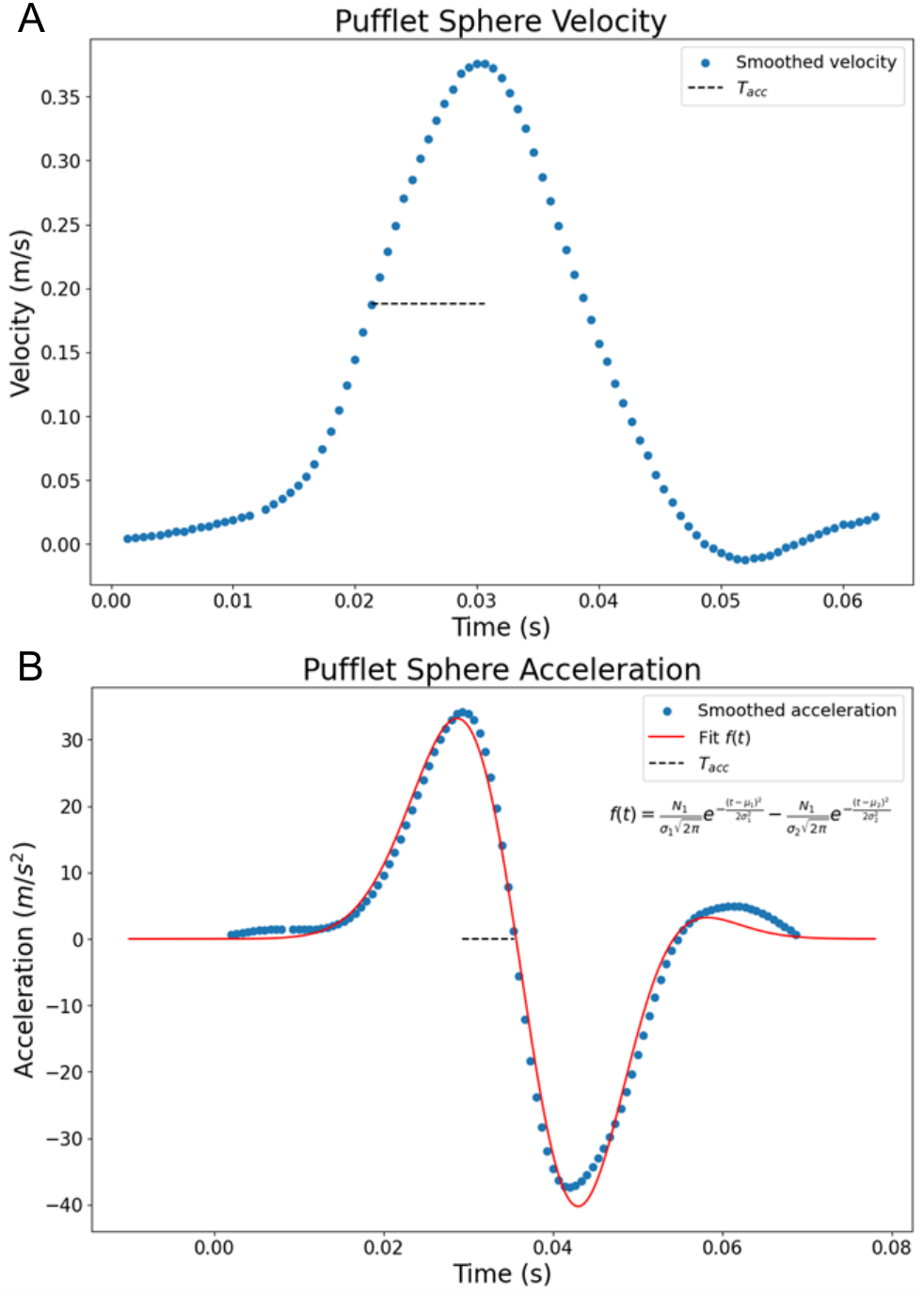}
    \caption{\textbf{(A)} Smoothed sphere velocity for a single Pufflet with $T_{acc}$ as the time between the sphere's half-max and max velocities. \textbf{(B)} Smoothed sphere acceleration for a single Pufflet with $T_{acc}$ as the time between the max acceleration and $a=0$. A double-Gaussian fit is performed to calculate $U_0$ and $T_{acc}$.}
    \label{fig:Re_Calcs}
\end{figure}

\subsubsection*{Experimental Single Pufflet - Vortex Tracking}
To experimentally identify the location of the vortex, we perform a high degree fit to the $v_y$ data along the $x$-axis, as shown in Figure~\ref{fig2}E. Then we solve numerically for the roots of the fit to find the $x$-intercept, and repeat this process 10,000 times, while varying each data point within its uncertainty. The reported vortex location is the average, and the standard deviation is the uncertainty.

\subsubsection*{Understanding transport by metachronal waves}

\paragraph*{Stokeslets wave.}As discussed in the main text, the memory kernel in the Pufflet solution makes analytical solutions for trajectories difficult to obtain. To understand the transition from a regime where singularities interact only weakly (i) to interaction (ii) to overall transport (iii) (see main text Fig.~\ref{fig4}A), we turn to the case of a metachronal wave of Stokeslets along the $z$-axis, separated by a distance $\delta$ and actuated one after the other. These solutions provide a foothold to make the calculations for the wave of Pufflets more tractable. We plot the same grid deformation as in Fig.~\ref{fig4}A for the case of a unit steady force active over a time interval $T$ such that $f T = 1$, and identify the same qualitative behaviour of cooperation and transport as for the inertial case in Figure~\ref{figwaveSM1}.

\begin{figure}[h!]
 	\centering
\includegraphics[width=.6\textwidth]{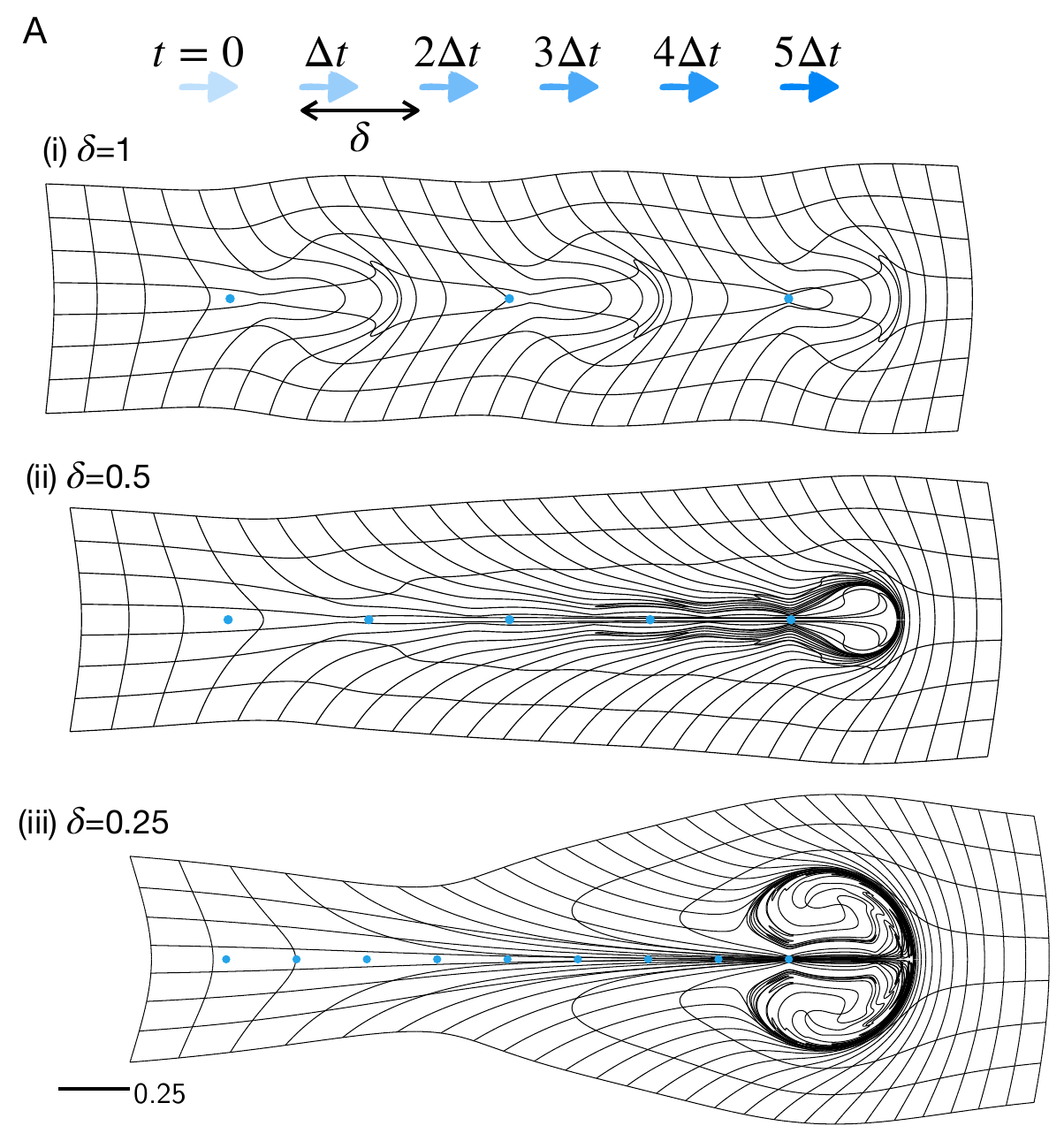}
 	\caption{Deformation of a square grid by a wave of Stokeslets with different spacings (see sketch), $\delta = 1, \, 0.5, \, \text{and} \, 0.25$.   }
    \label{figwaveSM1}
\end{figure}

We now move to a more scaling analysis of the transport by a wave. First, consider a point on the axis, at $(0,z_0)$. When a Stokeslet is active at $(0,0)$, the point moves according to  
\begin{equation}
    \frac{\dd z}{\dd t} = \frac{f}{4 \pi \mu} \frac{1}{\abs{{z}}}, 
\end{equation}
and we can use separation of variables to obtain the endpoint $z_f$ of a trajectory starting at $z_0$ 
\begin{equation}
    \begin{split}
        z_f \abs{z_f} &= z_0 \abs{z_0} + \frac{f T}{2 \pi \mu}, 
    \end{split}
\end{equation}
as plotted in Fig.~\ref{figwaveSM}A. 

In particular, the front of the mushroom-like structure generated by a single Stokeslet of constant strength $f$, active for a time interval $T$ at the origin, is therefore located at $z_M = \left(\frac{f T}{ 2 \pi \mu}\right)^{1/2}$. 
To study transport by a metachronal wave of such Stokeslets, we look for periodic points $z^*$ for which $z_f = \delta + z_0$. 
Analytically, we obtain the condition
\begin{equation}
    (z^*+\delta) \abs{z^*+ \delta} = z^* \abs{z^*} + \frac{f T}{2 \pi \mu}.
\end{equation}
On the plot of Fig.~\ref{figwaveSM}A, these correspond to the intersections $z^*$  with a horizontal line at $\delta$ if they exist. 
We distinguish three different cases: (1) the starting and arrival points are both behind the Stokeslet ($z_0 <0, \; z_f <0$), (2) the particle starts behind it and reaches the front ($z_0 <0, \; z_f > 0$) and (3) both are in front of the singularity ($z_0 >0, \; z_f >0$). 
\begin{equation}
\begin{split}
    \textrm{(1)} \,  - (z^*+\delta)^2 = - {z^*}^2 + \frac{f T}{2 \pi \mu}, \, \quad \textrm{so} \; \; \,   z^* = - \frac{f T}{ 4  \pi \mu \delta} + \frac{\delta}{2}   \;   \quad \quad \quad & \text{with} \quad \delta < \left( \frac{f T}{ 2 \pi \mu}\right)^{1/2}  \\
   \textrm{(2)} \quad \quad  (z^*+\delta)^2 = - {z^*}^2 + \frac{f T}{2 \pi \mu}, \;  \, \textrm{so} \quad   z^* = - \frac{\delta}{2} \pm \sqrt{\frac{f T}{ \pi \mu } - \delta^2}    \quad & \text{with} \quad \left(\frac{f T}{ 2 \pi \mu}\right)^{1/2}< \delta < \left( \frac{f T}{ \pi \mu}\right)^{1/2} \\
    \textrm{(3)} \quad  \quad (z^*+\delta)^2 = {z^*}^2 + \frac{f T}{2 \pi \mu}, \,  \quad \textrm{so}  \quad  z^* = \frac{f T}{ 4  \pi \mu \delta} - \frac{\delta}{2}  \; \; \quad \quad   \quad & \text{with} \quad \delta < \left(\frac{f T}{ 2 \pi \mu}\right)^{1/2}. 
\end{split}
\end{equation}
We validate these results numerically.

\begin{figure}[h!]
 	\centering
\includegraphics[width=.8\textwidth]{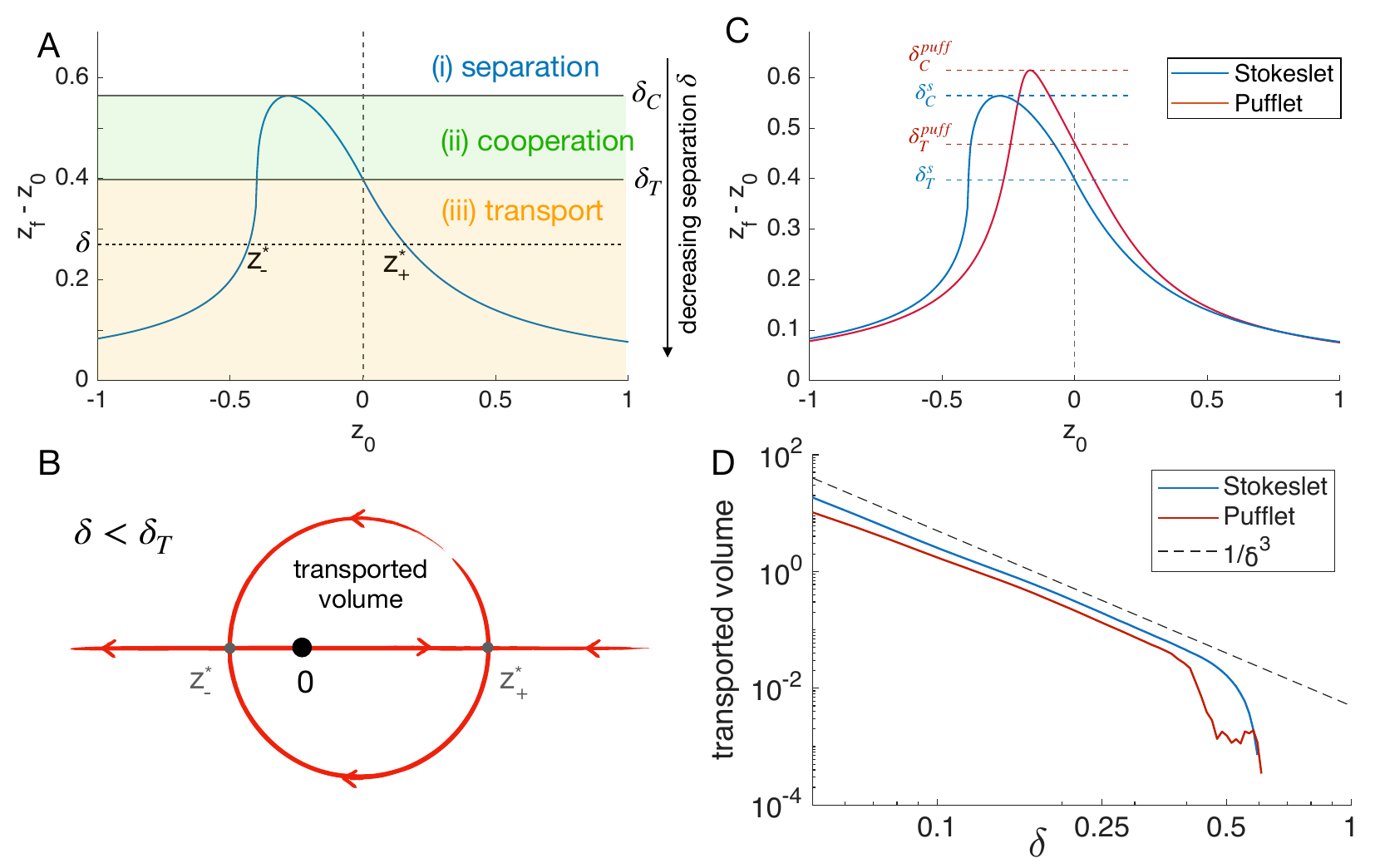}
 	\caption{ \textbf{(A)} Displacement by a single Stokeslet in the wave referential $z_f - z_0$ for a point starting at $z_0$. Fixed points for the metachronal wave are the points where $z_f-z_0 = \delta$  (horizontal dashed line). On this diagram, we identify the three regimes corresponding to (i) separated Stokeslets, (ii) cooperation, and (iii) transport. \textbf{(B)} Sketch of the fixed points and transported volume in the reference frame of the wave. \textbf{(C)} Comparison between the non-inertial (Stokeslet) and inertial (Pufflet) metachronal waves, showing that transitions to cooperation and transport, $\delta_C^{puff}$ and $\delta_T^{puff}$, happen earlier for Pufflets (dashed red lines) than for Stokeslets (dashed blue lines).  \textbf{(D)} Comparison between the transported volume by Pufflet and Stokeslet waves, both scaling as $\delta^{-3}$.   }
    \label{figwaveSM}
\end{figure}

The condition for cooperation is that some particles are transported for a distance greater than the Pufflet separation, which implies the existence of a fixed point. From the above analysis, this occurs below a critical separation $\delta_C =  (f T / (\pi \mu))^{1/2}$ (case (2) above). This corresponds to the transition from the regime of spatially distant Pufflets (i) to that of interacting Pufflets with no global transport (ii). 
However, this is not sufficient for an entire region of the fluid to be transported by the wave. From comparison with the simulations, we find that the second transition to the global transport regime (iii) coincides with the case where one of the fixed points is in front and one behind the point force (cases (1) and (3) above). This transition happens when the front of the first Stokeslet's mushroom exactly reaches the second singularity, so $z_M = \delta$, or equivalently, when $\mathbf{0}$ is a fixed point. Therefore, transport occurs when $\delta < \delta_T$, and $\delta_T = (f T / (2 \pi \mu))^{1/2} $.  We sketch in Fig.~\ref{figwaveSM}C the structure of the flow in this case. The distance between the front and back fixed points is $z_+^* - z_-^* = f T /(2 \pi \mu \delta) \sim \delta^{-1}$, which gives an intuitive explanation for the scaling of the transported volume $V \sim (z_+^* - z_-^*)^3 \sim  \delta^{-3} $ in Fig.~\ref{fig4}D and~\ref{figwaveSM}D.

\paragraph{Inertial and non-inertial metachronal waves.} We can now compare these regimes for non-inertial (Stokeslet) metachronal waves to the inertial case of a wave of Pufflets discussed in the main text. Because the Pufflet velocity field is time-dependent, separation of variables cannot be used, so we compare the results numerically. 
The three regimes are similar, but we find that inertia facilitates the transitions to cooperation and transport, which occur at lower spatial separations $\delta$, as shown by plotting the fixed points in Fig.~\ref{figwaveSM}C. On the other hand, the distance $z^*_+-z^*_-$ between the fixed points is reduced for a Pufflet, which mirrors the reduction in transported volume measured in Fig.~\ref{figwaveSM}D.

\subsubsection*{Other Supplementary Materials for this manuscript:}
Movies S1 to S7\\

\newpage










\clearpage 
\paragraph{Caption for Movie S1.}
\textbf{Lagrangian grid deformation in the flowfield generated by Pufflet}
Animated version of Figure~\ref{fig2}G. Simulation of Lagrangian grid (gray) displaced and
distorted by a single Pufflet. Chosen trajectories are plotted in color. Spacial dimensions are given in non-dimensional units.
\paragraph{Caption for Movie S2.}
\textbf{Lagrangian grid deformation in the flowfield generated by Cyclet}
Animated version of Figure~\ref{fig3}D. Simulation of Lagrangian grid (gray) displaced and
distorted by a single Cyclet - a pair of Pufflets in the same location firing with time delay. Chosen trajectories are plotted in color. Spacial dimensions are given in non-dimensional units.
\paragraph{Caption for Movie S3.}
\textbf{\textit{Spirostomum ambiguum} metachronal wave}
This video shows a \textit{Spirostomum ambiguum} cell with its cilia beating in metachronal waves. The liquid medium contains 1um tracer particles to demonstrate transport along a wave of Pufflets. The video was taken using Differential Interference Contrast (DIC) microscopy on a Nikon Eclipse Ti2 microscope. The magnification used was 20x, and the video was recorded at 190fps. The video plays at 1/19th speed. 
\paragraph{Caption for Movie S4.}
\textbf{Raw video for an experimental Pufflet}
This video is an example of raw video footage for a single Pufflet. The video was recorded using the setup described in the Experimental Methods section of the main text. The flow fields for the raw videos were extracted from using PIVlab software, and the flow fields were further analyzed using the \\\textit{\texttt{PIV\_9cm\_1500fps\_PaperCopy.ipynb}} notebook. Please note that this video is for a single Pufflet trial that was not used in the main text data. 
\paragraph{Caption for Movie S5.}
\textbf{Experimental Pufflet streamlines}
This video shows the streamlines and flow field magnitude from a single Pufflet trial. The video was generated using the \\\textit{\texttt{PIV\_9cm\_1500fps\_PaperCopy.ipynb}} notebook. 
\paragraph{Caption for Movie S6.}
\textbf{Experimental Cyclet tracer particle trajectories}
This video shows the particle trajectories found using TrackMate for a Cyclet overlaid on the original video used for the tracking. The trajectories in this video were used to calculate the normalized displacements shown in Figure 3C. 
\paragraph{Caption for Movie S7.}
\textbf{Wave of Pufflets}
Deformation of a square grid of fluid parcels by a wave of Pufflets over time, with the final deformation shown in Fig.4. The pufflet spacings are $\delta = 1, \, 0.5, \, \text{and} \, 0.25$ from top to bottom.




\end{document}